\documentclass{article}

\usepackage{amsmath}        

\usepackage{arxiv}

\usepackage[utf8]{inputenc} 
\usepackage[T1]{fontenc}    

\usepackage{hyperref}       
\usepackage{url}            
\hypersetup{hidelinks}
\usepackage{booktabs}       

\usepackage{amsfonts}       
\usepackage{mathabx}        

\usepackage{nicefrac}       
\usepackage{microtype}      
\usepackage{xcolor}         

\usepackage{graphicx}       
\usepackage{multirow}
\usepackage[toc,page, title, titletoc]{appendix} 

\usepackage{caption} 
\usepackage{cjhebrew} 

\usepackage[T1]{fontenc}
\usepackage[outline]{contour}
\contourlength{0.25pt}
\contournumber{1}

\usepackage{titlesec}
\titleformat{\paragraph}
{\normalfont\normalsize\bfseries}{\theparagraph}{1em}{}
\titlespacing*{\paragraph}
{0pt}{3.25ex plus 1ex minus .2ex}{1.5ex plus .2ex}

\usepackage{algorithm}
\usepackage[noend]{algpseudocode} 
\algnewcommand\algorithmicforeach{\textbf{for each}} 
\algdef{S}[FOR]{ForEach}[1]{\algorithmicforeach\ #1\ \algorithmicdo} 

\usepackage{subcaption}  

\newcommand{\medcup}{\mathbin{\scalebox{0.8}{\ensuremath{\bigcup}}}}  
\newcommand{\medcap}{\mathbin{\scalebox{0.8}{\ensuremath{\bigcap}}}}  

\usepackage[sorting=none, citestyle=numeric-comp]{biblatex}

\title{Co-Membership-based Generic Anomalous Communities Detection}
\author{
    Shay Lapid,\thanks{lapidshay@gmail.com} \space{}
    Dima Kagan,\thanks{dimakagan15@gmail.com} \space{}
    and Michael Fire\thanks{mickyfi@post.bgu.ac.il}\\
    Department of Software and Information Systems Engineering\\
    Ben-Gurion University of the Negev\\
}

\begin{document}
\maketitle

\begin{abstract}
Nowadays, detecting anomalous communities in networks is an essential task in research, as it helps discover insights into community-structured networks. Most of the existing methods leverage either information regarding attributes of vertices or the topological structure of communities. In this study, we introduce the Co-Membership-based Generic Anomalous Communities Detection Algorithm (referred as to \textit{CMMAC}), a novel and generic method that utilizes the information of vertices co-membership in multiple communities. \textit{CMMAC} is domain-free and almost unaffected by communities' sizes and densities. Specifically, we train a classifier to predict the probability of each vertex in a community being a member of the community. We then rank the communities by the aggregated membership probabilities of each community’s vertices. The lowest-ranked communities are considered to be anomalous. Furthermore, we present an algorithm for generating a community-structured random network enabling the infusion of anomalous communities to facilitate research in the field. We utilized it to generate two datasets, composed of thousands of labeled anomaly-infused networks, and published them. We experimented extensively on thousands of simulated, and real-world networks, infused with artificial anomalies. \textit{CMMAC} outperformed other existing methods in a range of settings. Additionally, we demonstrated that \textit{CMMAC} can identify abnormal communities in real-world unlabeled networks in different domains, such as Reddit and Wikipedia.

\end{abstract}


\section{Introduction}
We live in a networked world where almost everything can be represented as a network, including online social networks (OSNs)~\cite{Boyd2007}, a virus outbreak~\cite{EbolaOutbreakGraphSaey2015}, food recipes~\cite{IngredientNetworkTeng2012}, and a city’s water supply system~\cite{WaterNetworkNardo2013}. 
One attribute of complex networks is the formation of communities~\cite{Girvan2002}, which are clusters of relatively densely connected vertices with respect to the rest of the network~\cite{Girvan2002}. For instance, a group of OSN users who share a common subject of interest~\cite{TwitterCommonInterestsLim2012}, a team of coworkers, exposing each other to virus transmission~\cite{CoronaSNAGoncalve2020}, a family of ingredients from a certain cuisine~\cite{FoodPairingCusinesAhn2011}, or a city neighborhood corresponding to its water supply system~\cite{WaterNetworkClustersBui2020}. Analyzing such a community-structured network can help us gain meaningful insights into the objects represented by communities. 
For example, the detection of OSN communities promoting violent extremism and radicalization~\cite{OSNExtrimistCommsBenigni2017}, finding a group with a high potential to cause a pandemic outbreak~\cite{EpidemicOutbreakCommStructWang2011}, recommending recipe-enhancing ingredients substitution~\cite{IngredientNetworkTeng2012}, and locating a neighborhood likely to suffer from a water supply breakdown~\cite{WaterSystemClusteringPerelman2011}.

The ability to detect anomalous communities is crucial to the deduction of insights from community-structured networks. Such insights could help humanity cease a pandemic by an early reveal of a virus' hot spots~\cite{PandemicStopGeraghty2020}, identify groups of fake profiles spreading fake news~\cite{FightingFakeNewsCampan2017}, or prevent targeted violence toward minorities by uncovering hatred-inciting communities in an online social network~\cite{HateSocialMediaLaub2019}.

Over the last two decades, both industry and academic researchers proposed various solutions to address the problem of anomaly detection in networks~\cite{Kagan2017, Noble2003, Jimeng2005, Papadimitriou2010, Akoglu2010, Fire2012}, aiming to utilize them to gain insights into the analyzed networks. Altshuler et al.~\cite{Altshuler2013} demonstrated that by applying an anomaly detection algorithm on call recording logs of a country’s mobile network, they could classify events appearing in a particular time period as emergencies or not. While the majority of the conducted studies are mainly focused on uncovering anomalous vertices, only a handful focus on detecting anomalous communities~\cite{Singh2011, Miller2014, Gupta2014, Bridges2015, Zhao2016, Perozzi2018, BansalAnomComm2020, Luan2021}.

Most of these methods fail in scenarios where the anomalous communities are concealed properly in the background network, having similar properties as the rest of the communities.
For example, methods based on density and fraction of cross-boundary edges, such as \textit{Conductance}~\cite{Gleich2006Conductance}, tend to achieve poor results as the anomalous communities become either more sparse or harder to separate from other communities.

In this study, we introduce a novel generic network analysis and machine-learning-based algorithm to detect anomalous communities in complex networks. Motivated by Kagan et al.~\cite{Kagan2017}, demonstrating anomalous vertices could be determined by the number of improbable edges they have, we have hypothesized a community composed of many unexpected vertices, has a higher chance of being anomalous.

The approach we employ to test our hypothesis undergoes a classification problem. We predict the affiliation probability of vertices to their communities, and aggregate the resulting probabilities to determine the ``normality’’ of each community. We adopt a novel concept of utilizing the information of vertices' co-membership between communities, by formulating the aforementioned classification problem as a link-prediction problem in a new utility network, in which vertices are connected to community-representing vertices if they belong to the corresponding communities in the original network. More simply stated, we create a new network encapsulating the information of co-membership and then applies an anomaly detection algorithm to it.

The two significant advantages of our method are: (a) It utilizes only information of vertices’ co-membership to communities, which makes it agnostic to the density of communities. It can only be affected by the fraction of cross-boundary edges they contain, and specifically, to improve as it increases; and (b) the co-membership information is converted to a network, of which only its structural features are extracted and utilized, which makes it independent of a specific domain. Our method succeeds in cases when communities are hard to separate from the background. Additionally, it is generic, unlike most of the studies in this field, which rely on attributes of vertices (see Section~\ref{subsec:RelatedWorkAnomalousCommunities}). 

To evaluate our algorithm, we utilized two types of labeled datasets (see Section~\ref{subSection:NetworkDatasets}): (a) \textit{Fully simulated random generated networks}, infused with anomalous communities; and (b) \textit{real-world networks} infused with anomalous communities, that is, forcibly connecting randomly generated anomalous communities to other real communities in the real-world data. Additionally, we applied our method to two unlabeled real-world networks collected from Reddit\footnote{\url{https://www.reddit.com}} and Hebrew Wikipedia\footnote{\url{https://he.wikipedia.org}} revisions information. The results demonstrate that our method successfully identifies anomalous communities in all cases. In the first two cases, where simulated data or real-data perturbation are involved, causing them to be labeled datasets, our algorithm outperformed the baselines when the anomalous communities were sparse, small, and contained many cross-boundary connections. In the first unlabeled case, Reddit, our algorithm was able to identify two communities (subreddits) that presented peculiar activity, such as an utter failure of collaboration. In Wikipedia, where we depict an article as a community and the Wikipedians (editors) who edit it as vertices, our algorithm uncovered articles tainted with political agenda or violent content due to trolling.

The key contributions of our study are threefold:
\begin{itemize}
\item We have developed a novel generic algorithm for anomalous communities detection in complex networks.
\item We have developed a novel random network generation algorithm, which generates a random community-structured network, infused with anomalous communities. The algorithm includes the ability to control the number, size, and type (network generating algorithm) of the normal communities and the infused anomalous communities. The algorithm is well-suited to conduct research in the field of anomalous community detection in networks.
\item We have developed an open-source code of this study's framework to facilitate future research in the field of anomalous community detection in complex networks.
\end{itemize}

The remainder of this paper is organized as follows: Section~\ref{sec:RelatedWork} provides a brief overview of previous studies on anomalous vertices detection in networks, and on anomalous subgraphs and communities detection in networks. In Section~\ref{sec:Methods}, we describe in detail our anomalous community detection method and our anomaly-infused community-structured random network generator. In Section~\ref{Section:ExperimentalSetup} we describe the collection and creation of the networks’ datasets used to evaluate our algorithm, and the experiments performed to evaluate it. In Section~\ref{sec:Results}, we report our study’s results. In Section~\ref{sec:Discussion}, we analyze the evaluation results and insights and discuss our algorithm's limitations. Lastly, in Section~\ref{sec:ConclusionFutureWork}, we present our conclusion and offer future research directions. 

\section{Related Work}
\label{sec:RelatedWork}
Detecting anomalies in network-based data is an essential task for countless applications and areas~\cite{Leman2015}. In recent times, the ability to detect anomalous communities in networks, rather than stand-alone anomalies, became a high-impact task as well~\cite{Miller2014}. The following section surveys studies on anomalous vertices detection and anomalous communities detection in networks.

\subsection{Anomalous Vertices Detection}
\label{subsec:RelatedWorkAnomalousVertices}
Research and technology in the field of anomaly detection in networks have significantly evolved in the last two decades~\cite{Leman2015}. In 2003, Noble and Cook~\cite{Noble2003} were one of the first to study anomaly detection in network-based data. Their method assumed infrequently occurring substructures indicate anomalous behavior, while normal substructures reoccur much more often. In 2005, Sun et al.~\cite{Jimeng2005} proposed a method for detecting anomalous vertices in bipartite networks, where they calculated relevance scores between vertices and aggregated the results to one score per vertex, where a low score indicated an anomaly. In 2010, Papadimitriou et al.~\cite{Papadimitriou2010} presented a method to identify anomalies in a web network by comparing consecutive pairs of snapshots of the network by calculating similarity measures between them, where scores that were too low or high indicated an anomaly. In the same year, Akoglu et al.~\cite{Akoglu2010} proposed OddBall, a feature-based method to detect outliers in weighted networks. They chose pairs of features extracted from a vertex’s egonet, whose patterns obey power-laws. Vertices with significant deviation from the patterns are considered outliers. In 2012, Fire et al.~\cite{Fire2012} presented a method for detecting fake profiles on online social networks based on anomalies in a fake user’s social structure, namely the topology of the network. The method followed the intuition a fake profile randomly connects to other users in the network. In 2018, Kagan et al.~\cite{Kagan2017} proposed a generic unsupervised algorithm able to detect anomalous vertices based on network topological features alone, with the idea a vertex with many improbable edges has a higher likelihood of being anomalous. In 2019, Ding et al.~\cite{AnomalyDetectDing2019} introduced an interactive deep-learning-based approach, which allows the system to proactively communicate with the end-user to mitigate the lack of labeled data and enhance the anomaly detection performance. Recently, Guti\'{e}rrez-G\'{o}mez et al.~\cite{AnomalyDetectGomez2020} proposed MADAN, a parallelized method to rank and localize outlier vertices within their contexts, at different scales of a network, where they utilize heat kernel to smoothen signals around vertices at each scale, and the remaining highly concentrated signals after smoothing point to anomalous vertices.

\subsection{Anomalous Subgraphs and Communities Detection}
\label{subsec:RelatedWorkAnomalousCommunities}

In recent years, due to the increase in volume and sophistication of cyber-threats~\cite{JangCybersecuritySurvey2014}, the ability to detect a group of entities whose linkage is abnormal regarding the other network’s edges, namely, the detection of anomalous communities, has become a necessity and a valuable field of research~\cite{Miller2014}.

In 2011, Singh et al.~\cite{Singh2011} were the first to address the problem of anomalous subgraph detection rather than single anomalous vertices detection, by utilizing an approach from the field of signal processing, by detecting signal in noise - to detect un-hinted anomalous subgraphs in a background network. More specifically, they applied sparse principal component analysis to the network’s modularity matrix and checked for substantial deviation in the results compared to the expected results of a random subgraph. The method was tested on a simulated network.

In 2014, Miller et al.~\cite{Miller2014} studied a similar direction by adopting another method from the field of signal processing and proposed several algorithms based on the spectral properties of the principal eigenspace of a network’s residuals matrix. The algorithms analyze the residuals, comparing the residuals of an observed random subgraph to its expected value to find outliers and were able to demonstrate the detection of small, highly anomalous subgraphs, in real-world networks.

In the same year, Gupta et al.~\cite{Gupta2014} proposed SODA. Given a set of queried subgraphs, which are a subset of a network, Gupta et al. classified them as anomalous or not, and constructed an attribute-based classifier utilizing linear programming methods, namely, SIMPLEX. They used the classifier to predict the edge existence probability between each pair of vertices in each queried subgraph. They gave each subgraph an ``outlireness'' score based on the number of unexpected existing edges and the number of missing expected edges. Subgraphs with ``outlireness'' scores above some threshold were classified as anomalous subgraphs. Gupta et al. have achieved a precision score of 0.881 on a real-world dataset infused with generated anomalous subgraphs. As they lacked a labeled real-world dataset, they manually checked the highest-ranked subgraphs and reported to find interesting outliers.

In 2015, Bridges et al.~\cite{Bridges2015} proposed GBTER, a generalized version of the BTER model~\cite{Seshadhri2011}, which is a generative model that simulates a real-world community-structured network, assuming each vertex belongs to a single community. Additionally, they used a method of computing the probability distribution of a network given a generative model, which could derive the probabilities of the network's subgraphs. Bridges et al. used the GBTER model to simulate a normal network and a network infused with an anomalous community, and compared the anomaly-infused network probability distribution to the expected distribution of the normal network. The comparison was used to detect the existence of an anomaly in the network, and by examining the probabilities of the subgraphs, they detected the anomalous subgraph. Bridges et al. have achieved an AUC score of 0.936 in an experiment conducted on a small simulated network~\cite{Bridges2015}. 

In 2016, Zhao et al.~\cite{Zhao2016} referred to the task of detecting an anomalous subgraph as an optimization problem that tries to find a subset of vertices that maximizes overall abnormalities. The main contribution of the research was the introduction of parallel computation of such a problem. In 2017, Kumar et al.~\cite{Kumar2017} studied sockpuppetry in discussion communities, where they discovered sockpuppets behave differently from benign users. One example they discovered was the more clustered ego-networks are, the more likely they are to interact with each other. In the same year, Zheng et al.~\cite{Zheng2017} presented ELSIEDET, a three-stage sybil detection scheme identifying ``elite'' sybil users participating in the campaigns. The elite sybil users are highly rated accounts utilized to generate trustworthy and realistic-looking reviews.

In 2018, Perozzi et al.~\cite{Perozzi2018} proposed AMEN, an algorithm for ranking communities by a proposed ``normality'' measure, based on communities' coherency, that is, internally consistent, and externally separated from their boundaries, based on attributes and topological structure means. Perozzi et al. tested their algorithm on several real-world datasets while introducing labeled anomalous communities by disordering the topological structure of chosen communities and changing their vertices' attributes assignment. They have achieved AUC scores ranging from 0.18 to 0.60 on the different datasets.

In 2020, Bansal et al.~\cite{BansalAnomComm2020} presented ADENMN, which treated attributed networks as multiplex networks by splitting a network into different network layers, where each layer represented the network created by a certain attribute of the original network. By assigning the same latter ''normality'' score to each community at each layer, and accumulating the ``layer activity''-weighted scores to uncover anomalous communities. They achieved MAP scores ranging from 0.22 to 0.51 on real-world datasets injected with anomalous communities created similarly to the latter.

Recently, Luan et al.~\cite{Luan2021} proposed RM-CNN, a convolutional neural network classifying whether a network contains an anomalous community given an expected degrees model. They supplied the model with the residuals produced by subtracting the expected adjacency matrix of a random network generated by a given random network generating algorithm with a certain set of parameters from the actual adjacency matrix. They evaluated their model by utilizing a dataset composed of simulated networks, where the anomaly-containing networks contain a dense community generated by Erdős–R\'enyii~\cite{RandomGraphsErdosRenyi1959} random network generating algorithm embedded in the background network and achieved AUC scores ranging from 0.89 to 1.00.

\section{Methods}
\label{sec:Methods}

In this study, we apply concepts from the domains of \textit{Graph Theory}, \textit{Complex-Networks Analysis}, and \textit{Supervised-Learning} as building blocks of \textit{CMMAC} algorithm, whose purpose is to uncover anomalous communities in complex networks. In the following subsections, we describe, in detail, the phases of our anomalous communities detection algorithm and our anomaly-infused community-structured random network generator.

\subsection{Anomalous Communities Detection Algorithm}
\label{subSection:Algorithm}

\textit{CMMAC} requires two preliminary steps, (1) Community-detection in the examined network, whose results are stored as a \textit{partition map},\footnote{A set of key-value pairs, where each key corresponds to a community, and the matching value is a list of the community's vertices.} and (2) Splitting the network into train and test sets that share common structural properties, to allow training on part of the network and detect anomalous communities on the rest of network. We separated the network by splitting the \textit{partition map} into two \textit{partition maps}, such that the two resulting sets do not share common communities but may share common vertices.

According to our hypothesis, the task of detecting anomalous communities in networks requires the following main steps: (a) We begin by utilizing the two input partition maps to construct two bipartite networks, where each is composed of a group of vertices that ``represent'' communities in the original network, ``regular'' vertices, and edges denoting a ``regular'' vertex belongs to a community in the original network (see Section~\ref{subSubSection:ConstructingBipartiteNetwork}), (b) we then extract topological features of the newly created network and utilize the train set topological features to train a link-prediction classifier (see Section~\ref{subSubSection:ConstructingLinkPredictionClassifier}), and (c) lastly, based on the aggregation of the link-prediction classifier probability results of the test set, we extract meta-features and rank them. As anomalous communities tend to contain an improbable set of vertices, the corresponding anomalous community-representing vertices are more likely found at the bottom margin of the ranked meta-features (see Section~\ref{subSubSection:DetectingAnomalousCommunities}). In the following subsections, we will elaborate on each one of these steps. 

\subsubsection{Constructing a Bipartite Network}
\label{subSubSection:ConstructingBipartiteNetwork}

\begin{figure}[H]
    \centering
    \includegraphics[width=8cm]{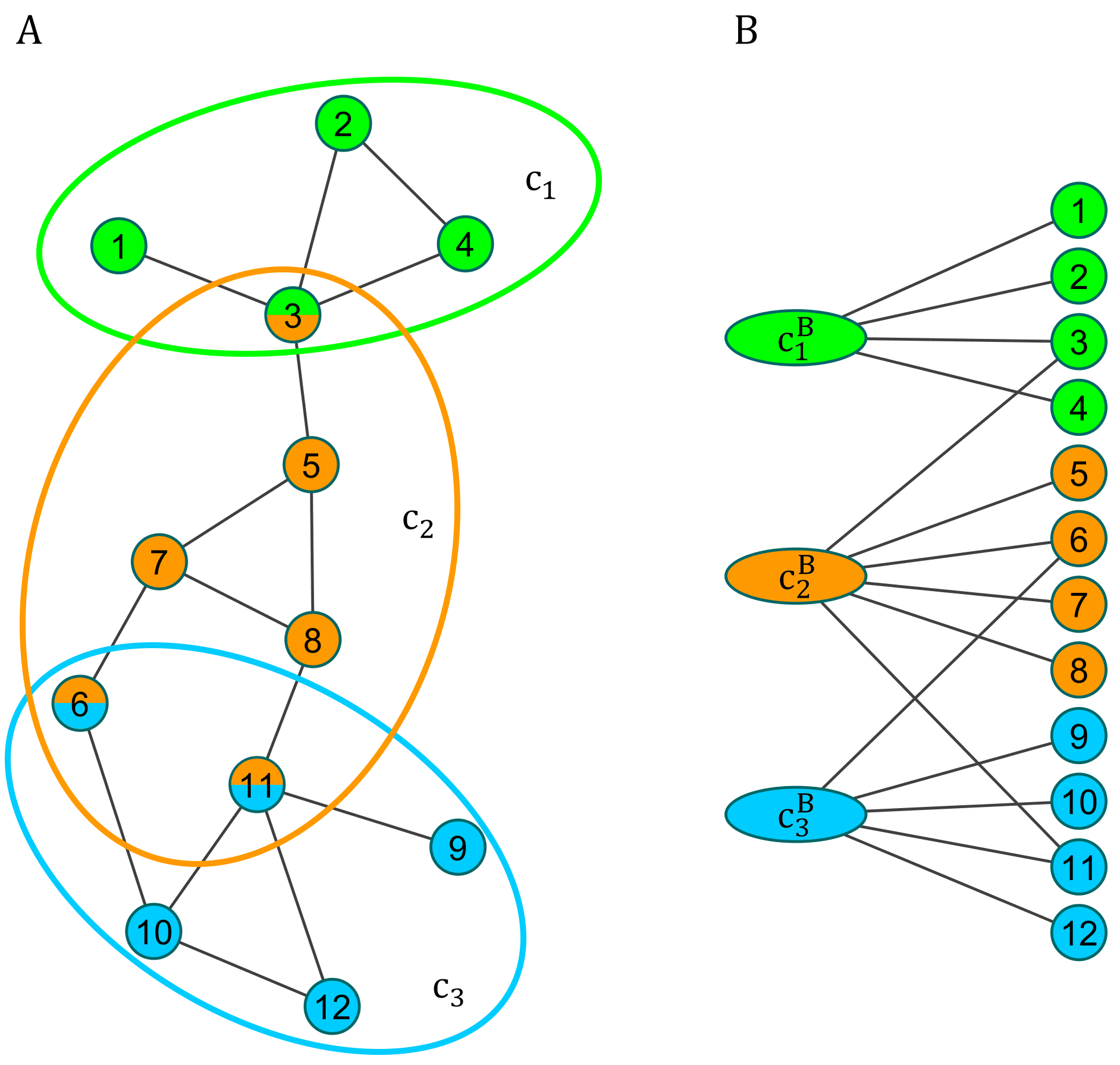}
    \caption{
        (A) an example of an overlapping-community-structured network $G$, where the set of vertices is $\{1,2,3,4,5,6,7,8,9,10,11,12\}\in V$, the set of communities is $\{c_{1},c_{2},c_{3}\}\in C$, and $c_1=\{1,2,3,4\}$, $c_2=\{3,5,6,7,8,11\}$, and $c_3=\{6,9,10,11,12\}$. (B) The corresponding derived bipartite network $BPG$ with the set of regular vertices $V$ and community-representing vertices $\{c^B_1, c^B_2, c^B_3\}\in C^B$, and the set of edges $E^B$ which state the belonging of a vertex to a community.
    }\label{fig:MethodsBipartiteFigure}
\end{figure}

To utilize information on vertices' co-membership between communities, we begin by creating two new bipartite undirected networks based on the given partition maps. Let $G := <V,E>$ be a network, where $V$ is a set of the network’s vertices, and $E$ is a set of the network’s edges, and let $\{c_{i}\}_{i=1}^n\in C$ be the set of $n$ communities in $G$.
We define the new bipartite network in the following manner: Using $G$,
we constructed a new bipartite undirected network $BPG := <V\medcup C^B ,E^B>$, where \(C^B:=\big\{{c_{i=1}^n}^B\in C^B|\forall{c^B_{i}}\in{C^B}, \exists{c_{i}}\in{C}\big\}\), and
$E^B:=\bigcup_{i=1}^nE_i^B$ where $E_i^B:=\big\{{(c_i^B,v_j)}|v_j\in c_i \: and\: c_i\in C\big\}$. Namely, using $G$. we constructed a new bipartite network in which the one part is composed of all the vertices $v\in V$ of $G$, and the other part consists of new vertices, where each vertex $c^B_i\in C^B$ in $BPG$ represents a community $c_i\in C$ in $G$. The undirected edges between a community-representing vertex and a regular vertex in $BPG$ stand for the belonging of the regular vertices to the corresponding communities in network $G$ (see Figure~\ref{fig:MethodsBipartiteFigure}).

\subsubsection{Constructing a Link-Prediction Classifier}
\label{subSubSection:ConstructingLinkPredictionClassifier}

After generating the communities' bipartite network, the next step of our algorithm is to construct a link-prediction classifier. The classifier's task is to produce a probability of the existence of an edge $(v,c^B)$ in $BPG$, given two vertices $v\in V$ and $c^B\in C^B$.

\paragraph{Feature Extraction}
\label{subSubSecTitle:FeatureExtraction}

Similar to Fire et al.~\cite{Fire2011EfficientTopologic}, we calculate a set of topological features for the edges to construct the link prediction classifier. We used only features which are meaningful for bipartite networks and modified them to adapt for analyzing bipartite undirected networks. Namely, we define the following features:

\begin{itemize}

    \item Let be $u_i\in \{V\medcup C^B\}$, a \textit{neighborhood} $\Gamma(u_i)$ is defined as the set of vertex $u_i$’s adjacent vertices:
    \[\Gamma(u_i):=\{u_j|(u_i,u_j)\in E^B\}\]
     
     The bipartite network reasons the following property - a \textit{neighborhood} of a vertex $v\in V$ only contains community-representing vertices $c^B\in C^B$ and vice versa.

    \item The \textit{degree} of $u_i \in \{V\medcup C^B\}$ is defined as:
    \[d(u_i):=|\Gamma(u_i)|\]

    \item For two vertices $v\in V$ and $c^B\in C^B$, the \textit{Total Friends} of $u$ and $v$ is defined as the number of distinct friends that $v$ and $c^B$ have together:
    \[TF(v,c^B):=|\Gamma(v)\medcup \Gamma(c^B)|\]
    
    \item As described in Section~\ref{subSection:AnomalyInfusedCommunityStructuredRandomNetwork Generator}, the \textit{Preferential Attachment Score} feature is based on the phenomenon that ``rich'' vertices increase their connectivity at the expense of the ``poor'' vertices~\cite{Barabasi1999}. We estimate how ``rich'' the two vertices $v\in V$ and $c^B\in C^B$ are, by multiplying  their degrees:
    \[PA(v,c^B) := |\Gamma(v)|\cdot|\Gamma(c^B)|\]
    
    \item The \textit{Friends Measure} hints two vertices connection ``strength'' by the number of connections between two vertices $v\in V$ and $c^B\in C^B$ \textit{neighborhood}s and is defined as:
    \[FM(v,c^B):=\sum_{x\in \Gamma(v)}\sum_{y\in \Gamma(c^B)} \delta(x,y)\]
    Where \(\delta(x,y)\) is defined as:
    \[\delta(x,y):=
    \begin{cases}
        1 & if \: x=y \: or \: (x,y)\in E^B \\
        0 & otherwise
    \end{cases}\]
    
    \item The \textit{Shortest Path} was demonstrated as a significant feature in the link-prediction task~\cite{Hasan2011Survey}. For two vertices \(v\in V\) and \(c^B\in C^B\) we define \textit{Shortest Path} as:
    \[SP(v,c^B):=
    \begin{cases}
        shortest\:path\:length\:between\:c\:and\:v^B\:in\:BPG & if\:a\:path\:exists \\
        -1 & otherwise
    \end{cases}\]

\end{itemize}

\paragraph{Classifier Construction}
\label{subSubSecTitle:ClassifierConstruction}

Similar to Kagan et al.~\cite{Kagan2017}, we train a link-prediction classifier on an equivalent number of positive and negative examples, where the edges taken into account are the train set bipartite network $BPG$ edges. We define a positive example as an existing edge $(v,c^B)\in E^B$, which stands for $v\in c$, or the belonging of vertex $v$ to community $c$ in the original network $G$. We define a negative example, as a non-existing edge $(v,c^B)\notin E^B$, which implies vertex $v\notin c$, namely, vertex $v$ does not belong to community $c$ in the original network $G$.

We uniformly sample positive and negative examples from the train network, and then calculate the features for each of the positive and negative edges in the train network and each edge in the test network. For each edge we calculate the edge features and the vertex features of both vertices (see Section~\ref{subSubSecTitle:FeatureExtraction}). Finally, we utilize the \textit{XGBoost} algorithm~\cite{Chen2016XGBoost} to construct the bipartite link-prediction classifier. We chose \textit{XGBoost} since previously conducted studies concluded \textit{XGBoost} performs well in terms of accuracy and efficiency in several cases of link-prediction tasks~\cite{Schlogl2019ComparisonAccidents, Mousa2018XGBoost}.\footnote{We evaluated \textit{XGBoost}, as well as other known effective link-prediction classifiers, such as \textit{Random Forest} ~\cite{Cukierski2011GraphBasedFeatures, Fire2011EfficientTopologic} and \textit{Feed-Forward Neural Network}~\cite{SandhyaPundhirUdayanGhose2020}. Our results indicated that the link prediction classifier constructed using the \textit{XGBoost} algorithm outperformed the other link prediction classifiers.}

\subsubsection{Detecting Anomalous Communities}
\label{subSubSection:DetectingAnomalousCommunities}

\begin{figure}[H]
    \centering
    \includegraphics[width=14cm]{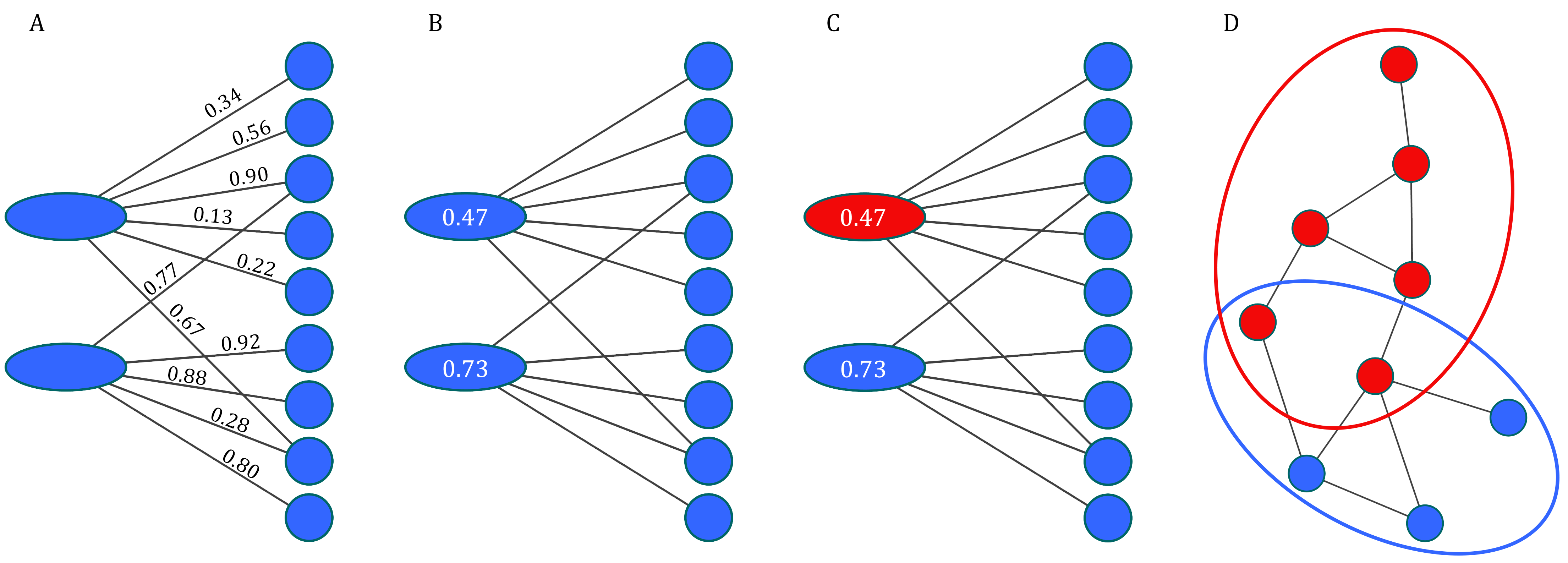}
    \caption{
        Algorithm overview. (A) After constructing the bipartite network, we created a link-predictor based on its topological features and predict edges' existence. (B) For each community-representing vertex, we aggregate the predicted probabilities into meta-features, for example, averaging them. (C) We rank them by the meta-features. (D) We fetch the corresponding original communities and manually examine them.
    }\label{fig:MethodsMetaFeatureAggregation}
\end{figure}

After constructing the link prediction classifier, we utilized it to create an unsupervised anomaly detection algorithm, which reduces the
complexity of searching for anomalies in a large space (see Figure~\ref{fig:MethodsMetaFeatureAggregation}). We utilized the link prediction classifier to emit the existence probabilities of all edges in the test network. Next, we aggregated the probabilities of the edges of community-representing vertices in several forms to create different meta-features. Then, we ranked the community-representing vertices by each one of the meta-features. Lastly, we manually examined the communities indicated by the community-representing vertices ranked at the bottom margins to find anomalous communities.

\paragraph{Meta-Feature Extraction}
\label{subSubSecTitle:MetaFeatureExtraction}

Inspired by Kagan et al.~\cite{Kagan2017}, we utilized the classifier to emit existence probabilities edges and aggregated them into meta-features. Based on the link-prediction classifier, we first provide formal definitions for the terms we use to describe the meta-features:
\begin{itemize}

    \item Let $p(v,c^B)$ be the probability of the existence of an edge $(v,c^B)$ in $BPG$ as emitted by the link-prediction classifier, where $v\in V$ and $c^B\in C^B$.
    
    \item Let $EdgeProbabilities(c^B):=\{p(v,c^B)|v\in \Gamma(c^B)\: and\: c^B\in C^B\}$ be the set of vertex $c^B$ edges’ existence probabilities.
    
    \item Let $EdgeLabels(c^B):=\{EdgeLabel(v,c^B)|v\in \Gamma(c^B)\: and\: c^B\in C^B\}$ be the set of vertex $c^B$ edges’ labels, that is, the label classifications of the edges with respect to a predefined \textit{threshold}, where $EdgeLabel(v,c^B)$ is defined as:
    \[EdgeLabel(v,c^B):=
    \begin{cases}
        1 & if \: p(v,c^B)\geq threshold \\
        0 & otherwise
    \end{cases}\]

\end{itemize}

Based on the above definitions, we define the following four meta-features as:

\begin{itemize}
    \item \textit{Edges Normality Probability Mean} is defined as the probability of a community-representing vertex $c^B$ to be normal, in other words, is the mean ($\mu$) taken over the existence probabilities of its edges:
    $$EdgesNormalityMean(c^B):=\mu(EdgeProbabilities(c^B))$$
    
    \item \textit{Edges Normality Probability STDV} is defined as one minus the standard deviation ($1 - \sigma$) of a set of vertex \(c^B\) edges’ existence probabilities:\footnote{STDV-based meta-features are preceded with $1-$ (one minus) since they behave the opposite from the rest of the features.}
    $$EdgesNormalitySTDV(c^B):= 1 - \sigma(EdgeProbabilities(c^B))$$
    
    \item \textit{Predicted Edge Labels Mean} is defined as the mean of the set of predicted labels of vertex \(c^B\)’s edges:
    $$PredictedEdgeLabelsMean(c^B):=\mu(EdgeLabels(c^B))$$
    
    \item \textit{Predicted Edge Labels STDV} is defined as the standard deviation of the set of predicted labels of vertex \(c^B\)’s edges:
    $$PredictedEdgeLabelsSTDV(c^B):= 1 - \sigma(EdgeLabels(c^B))$$

\end{itemize}

\paragraph{Meta-Feature Ranking}
\label{subSubSecTitle:MetaFeatureRanking}

After obtaining the meta-features of all community-representing vertices $c^B\in C^B$ in the test network, we ranked the vertices by each one of the meta-features. We then manually examined the communities indicated by the corresponding $k$ bottom vertices at each ranked meta-feature, where $k$ is a defined threshold.

\subsection{Anomaly-Infused Community-Structured Random Network Generator}
\label{subSection:AnomalyInfusedCommunityStructuredRandomNetwork Generator}
To evaluate the proposed method, we striven to generate community-structured networks similar to real-world scenarios. A mutual property of many complex networks is that the vertex connectivity follows a power-law distribution~\cite{RiseAndFallFire2019}; reflecting the fact new vertices attach preferentially to existing high-degree vertices~\cite{Barabasi1999}. Furthermore, in real-world networks, the majority of the communities have a certain extent of overlap~\cite{Ahn2010OverlappingCommunities, mcauley2012learning}, there exist vertices' co-memberships between them. 

Based on the above two statements, we reasoned generating a network where each community follows preferential attachment property and generating connections between these communities, would be a well-suited notion to mimic real-world overlapping community-structured networks. A simple implementation of the described concept would be to generate subnetworks using the Barab\'asi-Albert algorithm~\cite{Barabasi1999} and then connect them by connecting pairs of vertices from different subnetworks with a certain probability $p$.

We developed an algorithm encapsulating the essence of this concept and generalizing it further. Our algorithm creates subnetworks of two types, normal and anomalous, using two different random network generating algorithms. It then connects them in a ``\textit{dual-preferential}'' manner, by connecting vertices from ``new'' subnetworks to existing subnetworks by a probability corresponding to the subnetworks’ sizes, and within the chosen subnetworks to vertices with a probability corresponding to their degrees. The connected subnetworks are considered as overlapping communities in the created network. The two types of random network generating algorithms allow different types of ``normal'' and ``anomalous'' communities in the generated network.

The algorithm creates an overlapping community-structured network composed of normal and anomalous communities, given the following four parameters for each of the groups, normal communities and anomalous communities (see Algorithm~\ref{algorithm:Methods__NetworkGenerator}): (1) Random network generating algorithm (denoted $alg$), (2) a list of communities' sizes to create (denoted $comm\_sizes$), (3) arguments needed for random network generating algorithm (denoted $args$), and (4) a fraction of inter-connection to create between communities (denoted $inter\_p$). It returns the network, as well as its partition map describing the communities' belonging vertices. A detailed description of the algorithm is presented in Appendix~\ref{sec:Appendix__AnomalyInfusedCommunityStructuredRandomNetworkGenerator}, and we have published the implementation of the algorithm as an open-source code. 

\begin{algorithm}[H]
\caption{\newline Anomaly-Infused Community-Structured Random Network Generator}
\label{algorithm:Methods__NetworkGenerator}
\begin{algorithmic}[1]

\Procedure{GenerateRandomNetwork}{}
\Require $alg_{norm}, comm\_sizes_{norm}, args_{norm}, inter\_p_{norm}, alg_{anom}, comm\_sizes_{anom}, args_{anom}, inter\_p_{anom}$
\State $G\gets empty\: undirected\: network$
\State $Partitions\gets empty\: map$

\ForEach{$tuple$ in $[(alg_{norm}, comm\_sizes_{norm}, args_{norm}, inter\_p_{norm}),\newline \hspace*{5em}(alg_{anom}, comm\_sizes_{anom}, args_{anom}, inter\_p_{anom})]$}
    \State $alg, comm\_sizes, args, inter\_p \gets tuple$ \textcolor{gray}{\:\:\:// unpack tuple}
    \For{$c=1$ to $|comm\_sizes|$}
        \State $G^c=<V^c,E^c>\gets$ CreateNetwork$\big(alg(comm\_sizes[c], args)\big)$
        \State AddVerticesToNetwork($G, V^c$)
        \State AddEdgesToNetwork($G, E^c$)
        \State AddVerticesToPartition($c, V^c$)
    \EndFor
    \For{$c=1$ to $|comm\_sizes|$}
        \State InterConnect($V^c, inter\_p$)
    \EndFor
\EndFor
\State \textbf{return} $G, Partitions$
\EndProcedure
\item[]
\Procedure{InterConnect}{}
\Require $V^c, inter\_p$
    \For{$i=1$ to $\big\lfloor{|V^c|\times p \big\rfloor}$}
        \State $u\gets$ ChooseRandomVertex($V^c$)
        \State $otherComm\gets$ ChooseWeighted($comm\_sizes_{norm}$)
        \State $v\gets$ ChoosePreferentially($V^{otherComm}$)
        \State AddEdgesToNetwork($G, \{(u,v)\}$)
        \State AddVerticesToPartition($otherComm, \{v\}$)
    \EndFor
\EndProcedure

\end{algorithmic}
\end{algorithm}

\section{Experimental Setup}
\label{Section:ExperimentalSetup}

\subsection{Data Description}
\label{subSection:NetworkDatasets}

We evaluated our algorithm on two labeled datasets and performed two case studies by applying our algorithm on two unlabeled networks. We generated over 10 GB of networks data to be utilized throughout the study. The following subsections describe the creation of the datasets.

\subsubsection{Labeled Datasets}
\label{subsubsection:LabeledDatasets}

To the best of our knowledge, there are no publicly available network datasets with labeled anomalous communities. To evaluate our proposed algorithm we utilized two labeled datasets: (1) Networks created from Reddit subnetworks infused with anomalous communities, and (2) fully simulated networks with anomalous communities created by our network generator.\footnote{Refers to our Anomaly-Infused Community-Structured Random Network Generator throughout this section.} In the following subsections we describe the processes of acquiring real-world data, its perturbation to introduce labeled anomalous communities, and the generation of the fully simulated labeled networks.

\paragraph{Real-World Networks Infused with Artificial Anomalies}
\label{subSubSection:Datasets__RealWorldNetworksWithArtificialAnomalies}

Reddit is a popular collection of forums where people share news, content, or comment on others' posts. Reddit is composed of hundreds of thousands of communities, also called ``subreddits.'' Each subreddit is devoted to a different topic such as sports, sciences, and events~\cite{RedditWidman2020}. Using Reddit data, Jason Michael Baumgartner constructed a massive dump of Reddit comments, which he published and maintained~\cite{RedditCommentsBaumgartner2020}. This dataset contains the ID and the time each comment was posted, the subreddit it was posted in, the user who posted the comment, and the ID of the parent comment.\footnote{The comment ID to which the current comment replied.} In this study we utilized data obtained from the Reddit comments dataset, cleaned, and preprocessed by Fire et al.~\cite{RiseAndFallFire2019}.\footnote{\url{http://dynamics.cs.washington.edu/nobackup/reddit/reddit_last_graphs.tar.gz}.} The data contains over 2.37 billion posts posted from December 2005 through October 2016, by 19.72 million unique users, in 20,136 subreddits, each with more than 1,000 comments.

To evaluate \textit{CMMAC} on anomaly-infused real networks, we utilized the Reddit comments dataset to create 1,000 networks and infused them with generated anomalous communities for creating a dataset with ground truth labels. To create each of the anomaly-infused real networks, we sampled random subreddits from the Reddit comments dataset and constructed their networks. To follow the overlap property of real networks~\cite{mcauley2012learning} and to preserve a certain degree of co-membership information, which is required for \textit{CMMAC}, we constrained each sampled subreddit to have at least three users in common with at least two other subreddits in the network.

Formally, for each subreddit $s_i, i=1..k$, we define the subreddit’s network to be: $G^i:=<V^i, E^i>$, where $V^i$ is the set of vertices representing unique users who posted or commented within the subreddit $s_i$, and $E^i$ is the set of edges representing connections between users in subreddit $s_i$. Each edge $(u,v)\in{E^i}$ exists if a user $u$ replied to a comment, or as been replied to, by a user $v$, within subreddit $s_i$. For each subreddit $s_i, i=1..k$, exists at least two subreddits $s_m$ and $s_j$, where $|V_i \medcap V_m| \geq 3$ and $|V_i \medcap V_j| \geq 3$ . Next, we merged the $k$ networks into a single network, that is, $G:=<V, E>$, where $V=\medcup{V^k_{i=1}}$ and $E=\medcup{E^k_{i=1}}$.

\begin{table}[H]
    \centering
    \begin{tabular}{c|ccc}
        \toprule
        Sampled network &   $|V|$ & $|E|$ & avg. degree \\
        \midrule
        1 & 40,526 & 69,638 & 3.44 \\
        2 & 38,085 & 60,754 & 3.19 \\
        3 & 37,785 & 57,741 & 3.06 \\
        4 & 34,213 & 64,638 & 3.78 \\
        5 & 35,473 & 52,091 & 2.94
    \end{tabular}
    \caption{\label{table:Datasets__RedditNetowrksTable}Networks created by merging Reddit comments dataset subreddits' networks, each composed of 110 subreddits.}
\end{table}

Lastly, we utilized some functionality of our network generator, specifically, only the anomalous parameters tuple, (see Section~\ref{subSection:AnomalyInfusedCommunityStructuredRandomNetwork Generator}) to generate $a$ anomalous communities and attach them to the network. We attached them in a ``dual-preferential'' manner, by connecting vertices from the generated communities to subreddits chosen by a probability corresponding to their sizes, and within them to vertices chosen by a probability that corresponds to their degrees. Since the ``new'' vertex connected with a high chance to a ``central'' vertex in the subreddit it attached to, we considered it as part of the subreddit (A detailed  description of our network generator and the ``dual-preferential'' attachment property is presented in  Appendix~\ref{sec:Appendix__AnomalyInfusedCommunityStructuredRandomNetworkGenerator}). 

In this study, we constructed 1,000 networks as described above, by creating five networks composed of $k=110$ subreddits (see Table~\ref{table:Datasets__RedditNetowrksTable}), and attaching each of them 200 distinct sets of ten anomalous communities ($a=10$), where each generated with a different combination of parameters fed to our network generator.
The motivation for choosing the parameters is to get experiment results that properly present the regions where \textit{CMMAC} starts outperforming the other methods, and to learn its strengths and weaknesses. (For further details on the selection of network generation parameters see Appendix~\ref{sec:Appendix__SelectionOfNetworkGenerationParameters}). Specifically, we used the following parameters grid: (1) $alg_{anom}=$\textit{Erdős–R\'enyii}~\cite{RandomGraphsErdosRenyi1959}, (2) $args_{anom}\in{\{0.05, 0.1, 0.2, 0.4, 0.8\}}$,~\footnote{In \textit{Erdős–R\'enyii} random network generation algorithm $args_{anom}$ denotes $p$, the probability of edge existence between each pair of vertices.} (3) $inter\_p_{anom}\in{\{0.05, 0.1, 0.15, 0.2, 0.25, 0.3, 0.35, 0.4\}}$, and (4) $comm\_sizes_{anom}\in{\{q_{0}, q_{0.1}, q_{0.25}, q_{0.5}, random\}}$.\footnote{$comm\_sizes_{anom}$ is a list of 10 community sizes, sampled from the 110 normal community sizes distribution, where $q_{x}$ denotes quantile $x$ of the distribution and $random$ denotes uniform random sampling from the distribution.}

\paragraph{Fully Simulated Networks}
\label{subSubSection:Datasets__SimulatedNetwork}

Boshmaf et al.~\cite{SocialBotNetworkBoshmaf2011} studied the vulnerability of OSNs’ to large-scale infiltration by socialbots. They created a Socialbot Network (SbN), that is, a community of fake users that form many connections among each other to generate attraction from regular users. Next, the fake users randomly connect to real users in the targeted OSN. Then, to avoid detection due to anomalous structure, or due to the detection of one fake user who presented anomalous behavior, the SbN decomposes by deleting connections between the fake users. Finally, the SbN performs an attack of choice, usually information harvesting for spreading fake news~\cite{SocialBotNetworkBoshmaf2011}.

Inspired by Boshmaf et al.~\cite{SocialBotNetworkBoshmaf2011}, we utilized our network generator (see Section~\ref{subSection:AnomalyInfusedCommunityStructuredRandomNetwork Generator}) to evaluate \textit{CMMAC} on synthetic networks that simulate different points in the progress of the SbN decomposition and different networks’ properties, by generating 1,000 anomaly-infused community-structured random networks.

We chose parameters for the network generator so (1) the ``normal'' part of each network imitates the properties of a real network, in particular, the Reddit’s network described in Section~\ref{subSubSection:Datasets__RealWorldNetworksWithArtificialAnomalies}, and (2) the ``anomalous'' part will provide experimental results that properly present the regions where \textit{CMMAC} starts outperforming the other methods. For further details on the selection of network generation parameters see Appendix~\ref{sec:Appendix__SelectionOfNetworkGenerationParameters}.

We constructed the 1,000 fully simulated networks using the following parameters:
(1) $alg_{norm}=$\textit{Barab\'asi-Albert}~\cite{Barabasi1999},
(2) $args_{norm}=1$,~\footnote{In \textit{Barab\'asi-Albert} random network generation algorithm $args_{norm}$ denotes $m$, the number of edges that connect a new vertex to existing vertices.} 
(3) $inter\_p_{norm}=0.075$,
(4) $comm\_sizes_{norm}\in{\{random\_sample_i, i=1..5\}}$, where $random\_sample$is a set of 5 distinct lists, each composed of 110 community sizes, sampled from the Reddit comments dataset subreddit's sizes distribution,
(5) $alg_{anom}=$\textit{Erdős–R\'enyii}~\cite{RandomGraphsErdosRenyi1959},
(6) $args_{anom}\in{\{0.01, 0.02, 0.04, 0.08, 0.16\}}$ to compensate for the relatively low average degree of the normal communities,
(7) $inter\_p_{anom}\in{\{0.05, 0.1, 0.15, 0.2, 0.25, 0.3, 0.35, 0.4\}}$, and
(8) $comm\_sizes_{anom}\in{\{q_{0}, q_{0.1}, q_{0.25}, q_{0.5}, random\}}$ is a list of ten community sizes, sampled from the 110 normal community sizes distribution, where $q_{x}$ denotes quantile $x$ of the distribution and $random$ denotes uniform random sampling from the distribution.

\subsubsection{Unlabeled Real-World Networks}
\label{subSubSection:UnlabeledRealWorldNetworks}

This section describes the process of acquiring, cleaning, and preprocessing real-world unlabeled network datasets, to which we applied our algorithm to discover meaningful insights. We utilized the Reddit comments dataset, specifically data from the  ``r/Place'' project, and the Hebrew Wikipedia revisions data.

\paragraph{Reddit's r/Place Network}
\label{subSubSubSection:RedditPlaceNetwork}

On April $1^{st}$, 2017, a collaborative project and social experiment called ``\textit{r/Place}'' was initiated by Reddit~\cite{Cuthbertson2017}. The creators created a white $1000\times{1000}\:-\:$pixel canvas and posted it online with a call for users to edit it, hinting them for collaboration. Users could only change one pixel color every five minutes. After 72 hours, and more than a million unique users, the canvas colors had been changed more than 16.5 million times. The canvas turned into a beautiful skirmish of nations’ flags and symbols, ideologies, famous paintings and characters, and much more.

We utilized the Reddit comments dataset~\cite{RedditCommentsBaumgartner2020} by filtering 5.8 million comments from over one million unique users in 12,870 subreddits posted between April $1^{st}$ through April $4^{th}$ at midnight, particularly, at the time of the \textit{r/Place} project. We cleaned the data by removing comments that did not include information about the author. We then grouped the comments by subreddit and cleaned the data further by removing subreddits that contained less than 50 comments.

We processed the data further by filtering in only the $610$ subreddits that actively participated in the \textit{r/Place} project,\footnote{\url{https://draemm.li/various/place-atlas/}} and from those we chose $468$ subreddits that ranged in size from $50$ to $2,500$. We created a network from the resulting $468$ subreddits with the same process described by Fire et al.~\cite{RiseAndFallFire2019}. The formal definition is similar to the definition described in Section~\ref{subSubSection:Datasets__RealWorldNetworksWithArtificialAnomalies}, without overlapping constraints. The resulting network consisted of 181,019 vertices and 339,306 edges.

\paragraph{Hebrew Wikipedia Revisions Network}
\label{subSubSubSection:Datasets__HebrewWikipediaRevisionsNetwork}

Wikipedia is a free, open-content collaborative online encyclopedia maintained by volunteering editors, also called \textit{Wikipedians}. Wikipedia is one of the most popular websites~\cite{SimilarWeb2020}, containing more than 174 million articles, in more than 300 languages, and which are read monthly by 1.5 billion unique visitors as of November 2020~\cite{WikimediaStats2020}. In the last two years, the articles have been maintained by an average of 45 million edits per month, also called \textit{Revisions}, which are performed by an average of 70 thousand active \textit{Wikipedians}~\cite{WikimediaStats2020}.

We utilized Quarry,\footnote{\url{https://quarry.wmflabs.org/}} an online public interface for running \textit{SQL} queries against the Wikipedia database, to acquire all revisions made to articles in the Hebrew Wikipedia between January $1^{st}$, 2016, and July $14^{th}$, 2020 (To view the utilized query See Appendix~\ref{sec:Appendix__HebrewWikipediaSQL}). The data contains almost 7.5 million revisions, performed by 295,263 Wikipedians, in 269,355 articles. The revisions dataset contains the revision ID, as well as information about the Wikipedian who performed the revision, its timestamp, the ID of the parent revision,\footnote{The revision modified by the current revision.} and the article title that was revised. We inner-joined the data with itself on the $parent\_revision\_id$ attribute to get the network of revising Wikipedians answering to each other, where the articles they revise are considered as communities.

We further preprocessed the data by filtering in only articles with between $20$ and $80$ distinct revising users and users that revised between $5$ and $300$ unique articles, to enforce overlap between articles regarding revising users. The resulting data contained 72,633 revisions done in 2,123 articles.

Formally, for each article $a_i, i=1..k$, we define the article network to be: $G^i:=<V^i,E^i>$, where $V^i$ is the set of vertices representing unique Wikipedians who revised article $a_i$, and $E^i$ is the set of edges, representing the connections between Wikipedians within article $a_i$. Each edge $(u,v)\in{E^i}$ exists if Wikipedian $u$ revised Wikipedian $v$'s revision or the opposite, within article $a^i$. We then constructed the whole Hebrew Wikipedia revisions network by merging the article networks; that is, $G:=<V,E>$, where $V=\medcup{V^k_{i=1}}$ and $E=\medcup{E^k_{i=1}}$. The resulting network consisted of 12,736 vertices and 13,765 edges.

\subsection{Experiments}
\label{susubbSec:Evaluation}

To extensively evaluate our algorithm, we utilized the labeled datasets described in Section~\ref{subSection:NetworkDatasets}.
We first split the datasets to train and test sets as follows: For each of the labeled networks described in Section~\ref{subsubsection:LabeledDatasets}, we selected 20 communities for the train set to train a suitable link prediction classifier and 100 communities for the test set. Furthermore, we split the communities, so the train set was composed only of normal communities, while the test set was composed of 90 normal communities and ten anomalous communities (10\%), which represents an estimation of anomalies percentage in an average social network~\cite{Kagan2017}. For the $r/Place$ dataset described in Section~\ref{subSubSubSection:RedditPlaceNetwork}, we randomly chose $100$ communities for the train set and $350$ communities for the test set. Finally, for the Hebrew Wikipedia revisions dataset described in Section~\ref{subSubSubSection:Datasets__HebrewWikipediaRevisionsNetwork}, we randomly selected $100$ articles for the train set and $1,000$ articles for the test set.
For further details on our train-test split methodology, see Appendix~\ref{sec:Appendix__TrainTestSplitMethodology}.

For each of the labeled datasets, we began by analyzing the ranking predictive ability of each of the meta-features, namely, by ranking the communities by each of our algorithm’s meta-features, comparing the resulting rankings, and choosing one meta-feature to use for the comparison to other methods.
Then, we compared \textit{CMMAC}'s performance to other methods with respect to the three parameters ($comm\_sizes_{anom}$, $args_{anom}$, and $inter\_p_{anom}$) we used to create and attach the anomalous communities in our experiments.  

The other methods we utilize for the comparison are the following known topology-based measures and methods: (a) \textit{Average degree}~\cite{AvgDeg_Charikar2000} - the average degree of all vertices in a community; (b) \textit{Cut ratio}~\cite{CutRatio_Yang2012} - the fraction of existing cut edges out of all possible edges; (c) \textit{Conductance}~\cite{Conductance_Andersen2006} - the fraction of total edge volume that points outside the community); (d) \textit{Flake-ODF}~\cite{FlakeAvgODF_Flake2000} - the fraction of community vertices that have fewer edges pointing inside the community than to the outside; (e) \textit{Average-ODF}~\cite{FlakeAvgODF_Flake2000} - the average fraction of the community cut; (f) \textit{AMEN}~\cite{Perozzi2018}; and (g) \textit{ADENMN}~\cite{BansalAnomComm2020}.

To utilize the latter two algorithms in our study,\footnote{\textit{AMEN} was implemented in MATLAB, and was published as an open-source. \textit{ADENMN} implementation was never published as an open source.} we implemented in Python the \textit{Unattributed-AMEN/ADENMN} algorithm, which is based on \textit{AMEN}~\cite{Perozzi2018} and \textit{ADENMN}~\cite{BansalAnomComm2020} (both share the same topological-based part), but only considers the topological structure of the network and ignores the vertex attribute-based logic. Namely, it omits the vectors of attributes and corresponding weights, and the weights learning process. To be precise, \textit{Unattributed-AMEN/ADENMN} implements the ``normality score'' expressed by:
$$N=\sum_{\substack{i\in{C}, j\in{C}, \\ i\neq{j}}}\Big(A_{ij}-\dfrac{k_i\cdot{k_j}}{2\cdot{|E|}}\Big) - \sum_{\substack{i\in{C}, b\in{B}, \\ (i,b)\in{E}}}\Big(1-\min{\big(1, \dfrac{k_i\cdot{k_b}}{2\cdot{|E|}}}\big)\Big),$$
where $k_i$ denotes the degree of vertex $i$, $C$ denotes a community, $A$ denotes the adjacency matrix of $C$, $B$ denotes the set of boundary-vertices of $C$, and $E$ denotes the set of edges in the whole examined network.

Since our algorithm is a ranking algorithm, we utilize evaluation measures from the field of \textit{information retrieval}. Most of the baselines we compare to are simple, while we consider \textit{AMEN} and \textit{ADENMN} more advanced algorithms. To properly compare \textit{CMMAC} to them, we use the same evaluation measure they used~\cite{Perozzi2018, BansalAnomComm2020}, \textit{average precision} obtained from the \textit{AUC of the precision-recall curve}~\cite{Su2015PrecisionRecall}. To compare \textit{CMMAC} meta-features we use the measured \textit{MAP}, obtained by taking the mean of several \textit{average precision} scores.

To uncover anomalous communities in the unlabeled datasets, we first utilized \textit{CMMAC} to rank the communities in each of the networks’ test sets. Then, to reduce the problem searching space, we selected only the three communities ranked at the bottom by each meta-feature and intersected them into one set. We also utilized the described baselines to rank the communities and intersected their three bottom-ranked communities as well.

To report reliable results, since the data was unlabeled, we manually examined each of the resulting communities, both by \textit{CMMAC} and by the other methods, to seek anomalies: (1) in the Reddit \text{r/Place} project dataset, we comprehensively reviewed posts during and related to the \text{r/Place} project, within the examined subreddit, as well as posts that generally review the \text{r/Place} project and mentions the examined subreddits, and looked up anomalous behavior, and (2) in Wikipedia, we developed a code that produces a list of differences between each pair of consecutive revisions for a given article throughout a specific period. The output is composed of deletions and additions of content, as well as special actions, such as page protection activation. Finally, we sought anomalous behavior through extensive reviewing of the differences.

\section{Results}
\label{sec:Results}

The following section presents the results obtained from the experiments we conducted. First, we describe the evaluation results of the labeled datasets (see Section~\ref{subSection:Results__LabeledDatasets}). Then, we present the communities that were revealed when applying our method to two unlabeled real-world network use cases (see Section~\ref{subSection:Results__UnlabeledRealWorldNetworks}).

\subsection{Labeled Datasets}
\label{subSection:Results__LabeledDatasets}

\begin{figure}[H]
    \centering
    \textbf{Meta-Features Predictive Ranking Ability Procured in Reddit-Based Anomaly-Infused Networks Dataset}\par\medskip
    \begin{subfigure}[b]{0.6\textwidth}
        \centering
        \includegraphics[width=\textwidth]{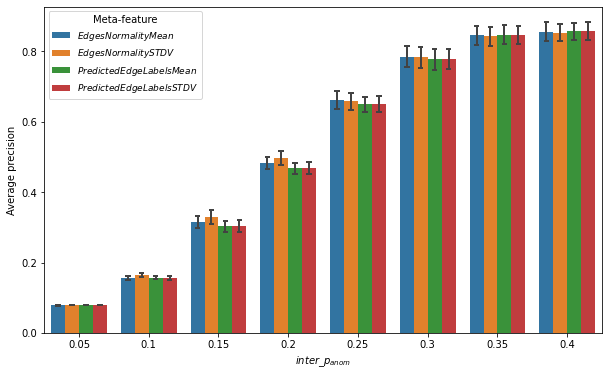}
        \caption{}
    \end{subfigure}
    \medskip
    \begin{subtable}[b]{0.3\textwidth}
        \centering
        \begin{tabular}{|l|c|}
            \hline
            Meta-features  &  MAP \\
            \hline
            $EdgesNormalityMean$    &   0.522   \\
            $EdgesNormalitySTDV$    &   \textbf{0.526}   \\
            $PredictedEdgeLabelsMean$   &   0.518   \\
            $PredictedEdgeLabelsSTDV$   &   0.518   \\
            \hline
        \end{tabular}
        \medskip\medskip\medskip\medskip\medskip
        \caption{}
    \end{subtable}
    \caption{(a) Average precision achieved by each meta-feature when applied to \textit{Reddit-based anomaly-infused networks}. Each bar color indicates a different meta-feature, and the whiskers indicate the confidence interval obtained by 125 networks' results. (b) MAP score of each meta-feature, calculated by taking the mean average precision scores at all $inter\_p_{anom}$ values range.}
    \label{fig:Results__RedditPredictiveAbilityDistribution}
    
\end{figure}

\begin{figure}[H]
    \centering
    \textbf{Meta-Features Predictive Ranking Ability Procured in Fully Simulated Networks Dataset}\par\medskip
    \begin{subfigure}[b]{0.6\textwidth}
        \centering
        \includegraphics[width=\textwidth]{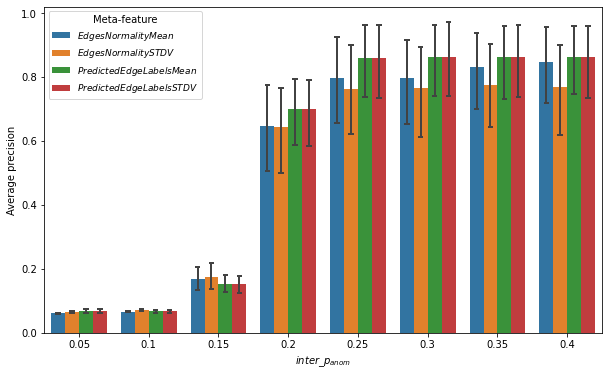}
        \caption{}
    \end{subfigure}
    \medskip
    \begin{subtable}[b]{0.3\textwidth}
        \centering
        \begin{tabular}{|l|c|}
            \hline
            Meta-features  &  MAP \\
            \hline
            $EdgesNormalityMean$    &   0.528   \\
            $EdgesNormalitySTDV$    &   0.504   \\
            $PredictedEdgeLabelsMean$   &   \textbf{0.554}   \\
            $PredictedEdgeLabelsSTDV$   &   \textbf{0.554}   \\
            \hline
        \end{tabular}
        \medskip\medskip\medskip\medskip\medskip
        \caption{}
    \end{subtable}
    \caption{(a) Average precision achieved by each meta-feature when applied to \textit{fully simulated networks}. Each bar color indicates a different meta-feature, and the whiskers indicate the confidence interval obtained by 125 networks' results.(b) MAP score of each meta-feature, calculated by taking the mean average precision scores at all $inter\_p_{anom}$ values range.}
    \label{fig:Results__SimulatedNetworkPredictiveAbilityDistribution}
    
\end{figure}

\begin{figure}[H]
    \centering
    \textbf{Reddit-Based Networks Dataset Evaluation and Comparison}\par\medskip
    \begin{subfigure}[b]{0.95\textwidth}
        \centering
        \includegraphics[width=\textwidth]{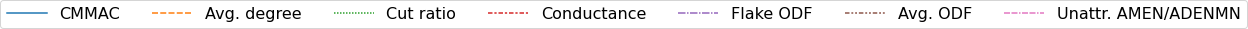}
    \end{subfigure}
    
    \begin{subfigure}[b]{1\textwidth}
        \includegraphics[width=16cm]{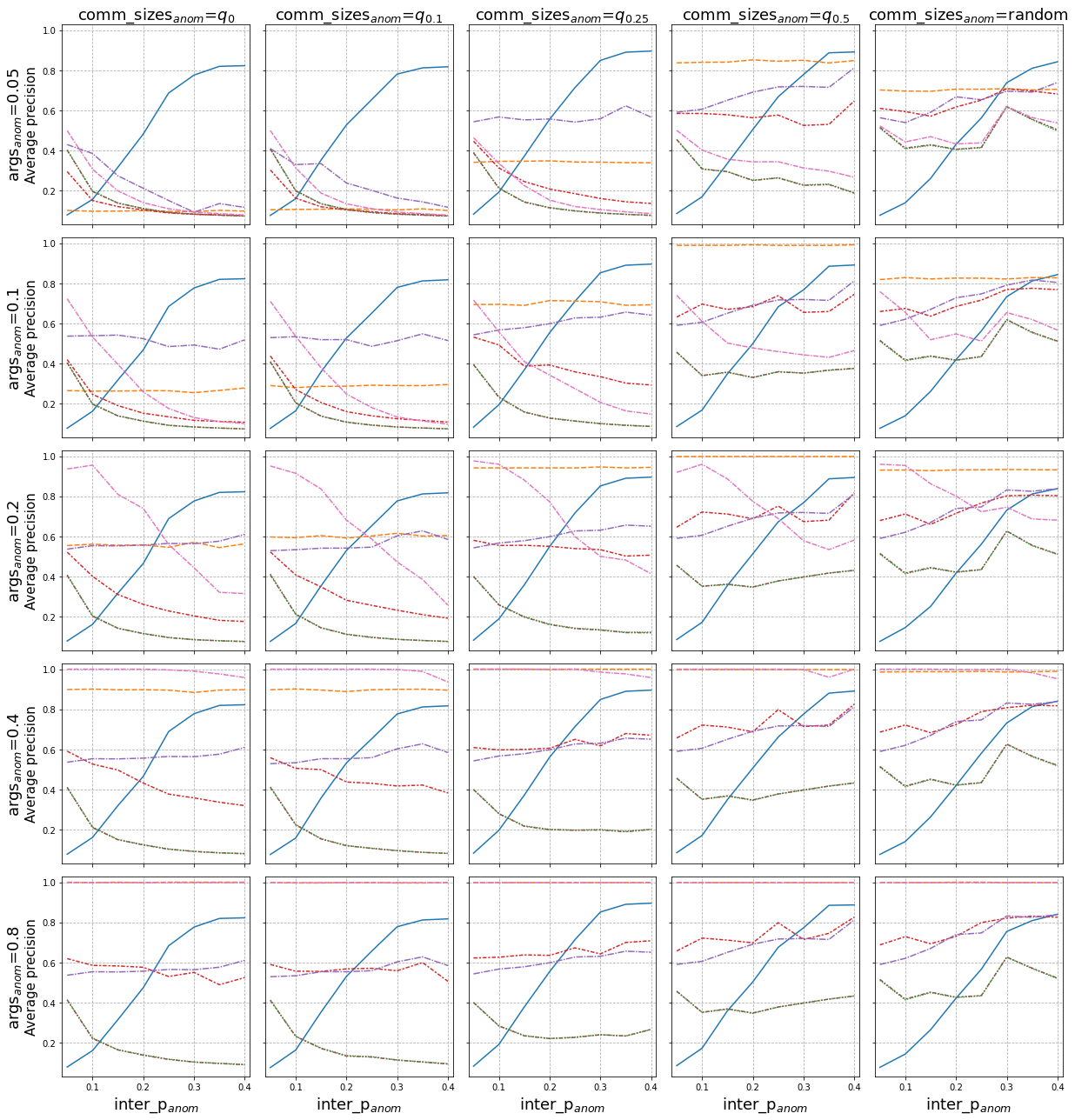}
    \end{subfigure}
    \caption{
        Evaluation of \textit{CMMAC}'s $EdgesNormalitySTDV$ meta-feature and comparison to other methods on \textit{Reddit-based anomaly-infused networks}. Each row of sub-plots describes a different $args_{anom}$ value, namely, the probability of edge existence between each pair of vertices in anomalous communities (density). Each column of subplots describes a different $comm\_sizes_{anom}$ value, that is, a quantifier describing the size of the anomalous communities as compared to the normal communities. At each of the subplots the X-axis describes $inter\_p_{anom}$ values, namely, the percentage of vertices in the anomalous communities that form ``dual-preferential'' inter-connections to normal communities. For each of the parameters obtained by this grid, we measured the \textit{average precision} score, obtained from the \textit{AUC of the precision-recall curve},~\cite{Su2015PrecisionRecall} which the Y-axis describes.
    }
    \label{fig:Results__EvaluationRedditWithArtificialAnomalies}
\end{figure}

\begin{figure}[H]
    \centering
    \textbf{Fully Simulated Networks Dataset Evaluation and Comparison}\par\medskip
    
    \begin{subfigure}[b]{0.95\textwidth}
        \centering
        \includegraphics[width=\textwidth]{Figures/Results__EvaluationHorizontalLegend.png}
    \end{subfigure}
    
    \begin{subfigure}[b]{1\textwidth}
        \includegraphics[width=\textwidth]{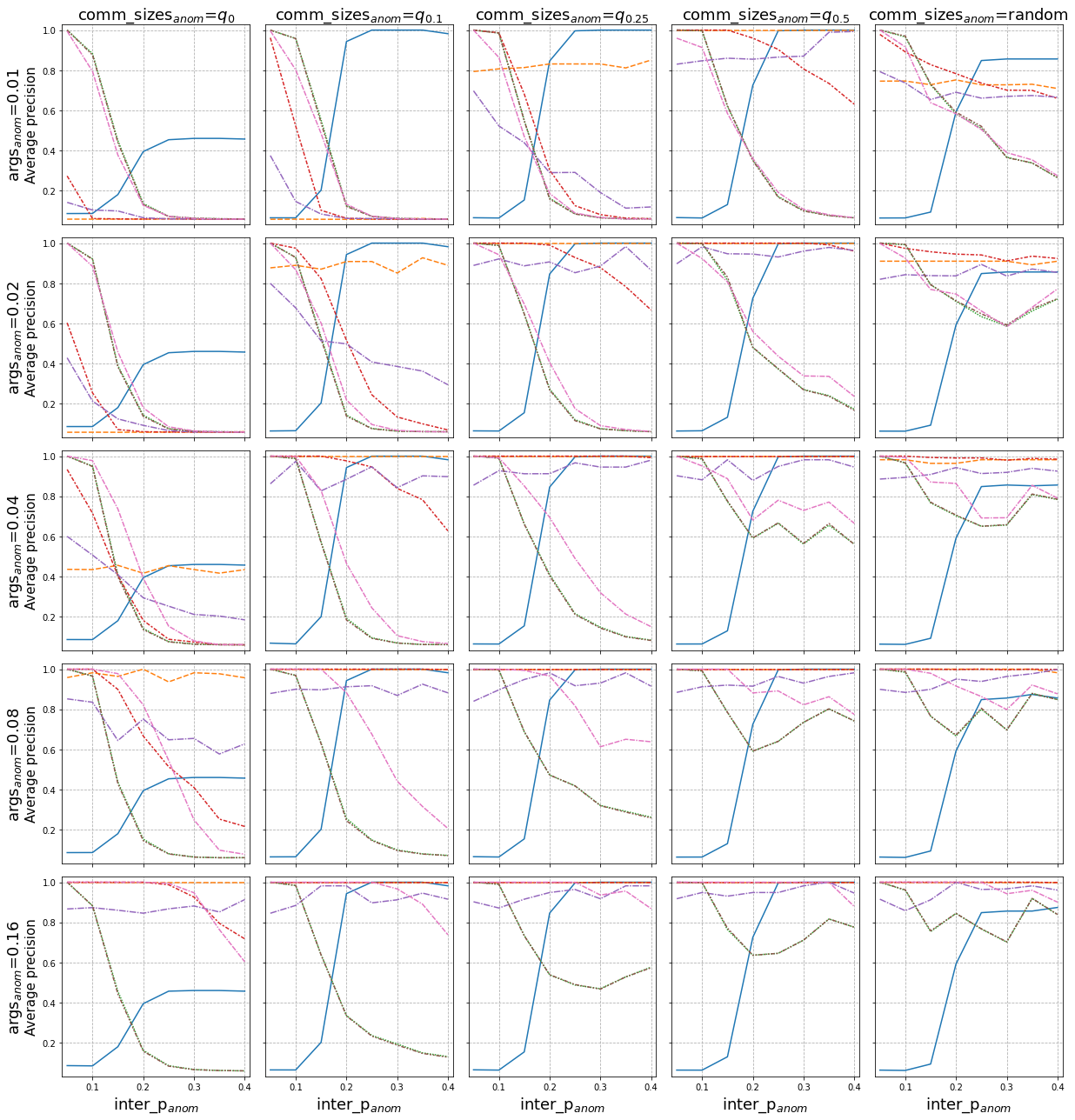}
    \end{subfigure}
    
    \caption{
       Evaluation of \textit{CMMAC}'s $PredictedEdgeLabelsSTDV$ meta-feature and comparison to other methods on \textit{fully simulated networks}. Each row of subplots describes a different $args_{anom}$ value, particularly the probability of edge existence between each pair of vertices in anomalous communities (density). Each column of subplots describes a different $comm\_sizes_{anom}$ value, that is, a quantifier describing the size of the anomalous communities as compared to the normal communities. At each of the subplots, the X-axis describes $inter\_p_{anom}$ values, namely, the percentage of vertices in the anomalous communities that form ``dual-preferential'' inter-connections to normal communities. For each of the parameters obtained by this grid, we measured the \textit{average precision} score, obtained from the \textit{AUC of the precision-recall curve},~\cite{Su2015PrecisionRecall} which the Y-axis describes.
    }
    \label{fig:Results__EvaluationSimulatedNetworkFigure}
\end{figure}

To evaluate our method on labeled datasets, we utilized the 2,000 networks we created in two datasets, as described in Section~\ref{subsubsection:LabeledDatasets}.
We first analyze the predictive ranking ability of the meta-features in each of the datasets.
Within the Reddit-based networks dataset, among all meta-features, the $EdgesNormalitySTDV$ achieved the highest MAP score of 0.526 (see Figure~\ref{fig:Results__RedditPredictiveAbilityDistribution}), while within the fully simulated networks dataset, the $PredictedEdgeLabelsMean$ and the $PredictedEdgeLabelsSTDV$ meta-features achieved the highest MAP score of 0.554 (see Figure~\ref{fig:Results__SimulatedNetworkPredictiveAbilityDistribution}).
Consequently, we utilized the latter meta-features to compare \textit{CMMAC} to the other methods. Specifically, we utilized the $EdgesNormalitySTDV$ meta-feature to report the comparison results in the Reddit-based networks dataset (see Figure~\ref{fig:Results__EvaluationRedditWithArtificialAnomalies}), and the $PredictedEdgeLabelsSTDV$ meta-feature to report the comparison results in the fully simulated networks dataset (see Figure~\ref{fig:Results__EvaluationSimulatedNetworkFigure}).


\subsection{Unlabeled Real-World Networks}
\label{subSection:Results__UnlabeledRealWorldNetworks}
This section presents the findings we discovered in two real-world unlabeled datasets. We utilized our algorithm to rank the communities by each of the meta-features. We then chose only the distinct communities ranked at the three lowest rankings by each of the meta-features (see subreddits list in 
Appendix~\ref{sec:AppendixDResultsUnlabeled}), thereby we reduced the searching space drastically. We then manually examined each of the resulting communities described in Section~\ref{susubbSec:Evaluation} and encountered interesting case studies. To verify the fidelity of the results obtained by \textit{CMMAC}, we also utilized all the other methods described in Section~\ref{susubbSec:Evaluation} and manually examined their results.

\subsubsection{Reddit's r/Place Network}
\label{subSubSubSection:ResultsRedditPlaceNetwork}

To evaluate \textit{CMMAC} on Reddit's \textit{r/Place} project test set we utilized the network construction method described in Section~\ref{subSubSubSection:RedditPlaceNetwork}. We utilized \textit{CMMAC} to rank the subreddits by each of the meta-features, as well as utilized the other methods to rank the subreddits. We selected the three bottom-ranked subreddits (see Tables~\ref{table:Appendix__TableRedditPlaceCMMACResults} and ~\ref{table:Appendix__TableRedditPlaceOtherMethodsResults}). Intersecting the three bottom-ranked subreddits resulted in ten distinct subreddits returned by \textit{CMMAC}, and nine distinct subreddits returned by the other methods, where there are no common subreddits between our algorithm and the other methods.

We manually examined all the subreddits as described in Section~\ref{susubbSec:Evaluation}, and exposed the following subreddits which presented abnormal behaviors, which were returned by \textit{CMMAC}: 

\begin{itemize}
    \item \textbf{\textit{r/BlueCorner}}\footnote{\url{https://www.reddit.com/r/BlueCorner/}} - According to the Redditor \textit{Andrewcshore315}~\cite{Andrewcshore2017TheBlueCorner}
    a member of the \textit{r/TheBlueCorner} subreddit, the \textit{r/BlueCorner} began as a violent subreddit that tried to paint the whole canvas with blue pixels, ruining other artifacts on its way.
    Due to its behavior, it quickly gained enemies, which made its users cease to cooperate and eventually abandon it. Many of the deserting users joined a new subreddit called \textit{r/TheBlueCorner}, which was led by new leadership, this time aiming to maintain the structure of the blue corner while respecting and protecting other arts.
    The subreddit \textit{r/BlueCorner} was ranked 349 out of 350, that is, $2^{nd}$ from the bottom by \textit{CMMAC}'s $PredictedEdgeLabelsSTDV$ meta-feature. For the rankings achieved by the other meta-features and by the other methods see Table~\ref{table:Results__UnlabledRankings}.

    \item \textbf{\textit{r/COMPLETEANARCHY}}\footnote{\url{https://www.reddit.com/r/COMPLETEANARCHY/}} - A comprehensive inspection involving a systematic investigation through its posts during and related to the \textit{r/Place} project, as well as questioning some of the participants brought up that subreddit \textit{r/COMPLETEANARCHY} presented an anomalous behavior - a complete failure of collaboration. Namely, there were many attempts to propose ideas, tactics, and courses of action, which were hardly commented upon and were never executed. Today there are no traces of their participation in the \textit{r/Place} project. The subreddit \textit{r/COMPLETEANARCHY} was ranked 348 out of 350, that is, $3^{rd}$ from the bottom by \textit{CMMAC}'s $PredictedEdgeLabelsSTDV$ meta-feature. For the rankings achieved by the other meta-features and by the other methods see Table~\ref{table:Results__UnlabledRankings}.
    
\end{itemize}
We could not detect any anomalous subreddits among the subreddits returned by the other methods.

\subsubsection{Hebrew Wikipedia Revisions Network}
\label{subSubSubSection:ResultsHebrewWikipediaRevisionsNetwork}

To evaluate \textit{CMMAC} on the Wikipedia revisions network test set, we utilized the network construction method described in Section~\ref{subSubSubSection:Datasets__HebrewWikipediaRevisionsNetwork}. We utilized \textit{CMMAC} to rank the article-representing communities by each of the meta-features, as well as utilized the other methods to rank the communities. We selected the three bottom-ranked articles (see Tables~\ref{table:Appendix__TableWikipediaCMMACResults} and ~\ref{table:Appendix__TableWikipediaOtherMethodsResults}). Intersecting the three bottom-ranked articles resulted in six distinct articles returned by \textit{CMMAC} and 12 distinct articles returned by the other methods. \textit{CMMAC}'s results and the other methods' results shared one common article.

We manually examined all the articles as described in Section~\ref{susubbSec:Evaluation}, and detected the following article, which presented anomalous behavior within its revisions history according to \textit{CMMAC} results:

\begin{itemize}
    \item \textbf{\textit{COVID-19 effects on the (Israeli) education system}}\footnote{Refers to the Hebrew Wikipedia page ``\<h/sp`t mgpt hqwrwnh `l m`rkt h.hjynwK>''.}~\cite{WikipediaCovidEducation} - The article \textit{COVID-19 effects on the education system}~\cite{WikipediaCovidEducation} was created in April 2020 and was dedicated to the effect of COVID-19 on the Israeli education system. In the two and a half months it existed within our dataset, many of its revisions were in regard to the gaps between the different Israeli society sectors, and government criticism. This article was used as a fertile ground for a political scuffle, due to its fast popularity gaining on account of COVID-19 related news article. The article \textit{COVID-19 effects on the education system} was ranked 998 out of 1,000 articles, that is, $3^{rd}$ from the bottom by \textit{CMMAC}'s $EdgesNormalitySTDV$ meta-feature. For the rankings achieved by the other meta-features and by the other methods see Table~\ref{table:Results__UnlabledRankings}.
\end{itemize}
We could not detect any anomalous articles among the articles returned by the other methods.

\begin{table}[H]
    \centering
    \begin{tabular}{ll|p{2cm}p{3.7cm}|p{3cm}}
         &  & \multicolumn{2}{c|}{Reddit's r/Place} & \multicolumn{1}{c}{Wikipedia Revisions}\\
         &  & \multicolumn{2}{c|}{(350 Subreddits)} & \multicolumn{1}{c}{(1,000 Articles)}\\
         &  & \textit{r/BlueCorner} & \textit{r/COMPLETEANARCHY}  & \textit{COVID-19 effects on the education system}  \\
        \midrule
        \multirow{4}{1em}{\rotatebox{90}{\textit{CMMAC}}} & $EdgesNormalityMean$ & 321 & 297 & 989 \\
        & $EdgesNormalitySTDV$ & 222 & 197 & \textbf{998} \\
        & $PredictedEdgeLabelsMean$ & 336 & 334 & 994 \\
        & $PredictedEdgeLabelsSTDV$ & \textbf{349} & \textbf{348} & 994 \\
        \midrule
        \multirow{6}{1em}{\rotatebox{90}{Other Methods}} & \textit{Average degree} & 106 & 195 & 952 \\
        & \textit{Cut ratio} & 80 & 166 & 663 \\
        & \textit{Conductance} & 79 & 151 & 639 \\
        & \textit{Flake-ODF} & 16 & 27 & 373 \\
        & \textit{Average-ODF} & 80 & 166 & 661 \\
        & \textit{Unattributed-AMEN/ADENMN} & 77 & 165 & 688\\
        \bottomrule
    \end{tabular}
    \caption{The ranking of the anomalous communities we ranked by each of \textit{CMMAC}'s meta-features and other methods.}
    \label{table:Results__UnlabledRankings}
\end{table}

\section{Discussion}
\label{sec:Discussion}

By analyzing the results presented in Section~\ref{sec:Results}, the following can be noted:

First, by analyzing the behavior of \textit{CMMAC} along the X-axis, namely the $inter\_p_{anom}$ values, at each of the subplots in Figures~\ref{fig:Results__EvaluationRedditWithArtificialAnomalies} and~\ref{fig:Results__EvaluationSimulatedNetworkFigure}, we can conclude that \textit{CMMAC} performance is correlated to the fraction of inter-connections between anomalous communities and other communities. Specifically, it performs better as the fraction of inter-connections arises. Namely, when more cross-boundary edges exist between anomalous and other communities. As the percentage of inter-connections increases, more vertices communities' co-membership information is available to \textit{CMMAC}, thereby enhancing its performance. However, the inferior results achieved by setting lower inter-connections fraction values indicate a limitation of \textit{CMMAC}. Namely, \textit{CMMAC} depends on a somewhat degree of overlap between communities in the examined network. Networks without overlapping communities lack essential information for \textit{CMMAC} to work properly. Nonetheless, most of the communities in real-world networks tend to overlap~\cite{Ahn2010OverlappingCommunities, mcauley2012learning}.

Second, \textit{CMMAC} requires inputs in form of partition maps that indicate each community's contained vertices. The creation of the partitions maps relies on a preliminary step of detecting overlapping communities in the observed network. The latter is a hard task, especially when utilizing only structural properties~\cite{OverlappingCommunityDetectionVieira2020}. The combination of the dependency on detecting the overlapping communities, and the fact overlapping between communities is essential for \textit{CMMAC}, presents a limitation of our approach. When we utilize non-network data and model it as a network in which we create communities according to the definition of the problem, we skip the need of detecting overlapping communities. For example, the creation of the Wikipedia revisions network is described in Section~\ref{subSubSubSection:Datasets__HebrewWikipediaRevisionsNetwork}.

Third, by analyzing the subplots along the rows in Figures~\ref{fig:Results__EvaluationRedditWithArtificialAnomalies} and~\ref{fig:Results__EvaluationSimulatedNetworkFigure}, namely, the densities of the anomalous communities, we can conclude \textit{CMMAC} is not affected by the density of a community, while all the other \textit{internal-consistency}-based\footnote{The degree of how community’s vertices are internally well connected.} methods are, videlicet, \textit{Average degree}, \textit{Conductance}, \textit{Flake-ODF}, and \textit{Unattributed-AMEN/ADENMN}. Specifically, all the \textit{internal-consistency}-based methods' performances degraded as the anomalous communities get sparser.

Fourth, by examining the evaluation results concerning the size of anomalous communities', i.e., along the columns in Figures~\ref{fig:Results__EvaluationRedditWithArtificialAnomalies} and~\ref{fig:Results__EvaluationSimulatedNetworkFigure}, we infer \textit{CMMAC} is not affected by the size of a community, whereas all the other methods are affected by the size. In particular, the other methods achieve poorer scores as the anomalous communities become smaller.

Fifth, according to the overall evaluation results in Figures~\ref{fig:Results__EvaluationRedditWithArtificialAnomalies} and~\ref{fig:Results__EvaluationSimulatedNetworkFigure}, we can conclude that \textit{CMMAC} outperforms other methods in the cases where the properties of the anomalous communities become similar to the rest of the communities, and when there are many cross-boundary edges between the anomalous communities and the other communities. Simply put, in the scenarios where the anomalous communities are small, sparse, and hard to separate from the other communities. It is important to keep in mind the latter finding was achieved and holds for a network whose structure follows a power-law distribution. To the best of our knowledge, no other method utilizes co-membership information. Particularly, the methods we utilized as baselines are founded upon either \textit{internal-consistency}, \textit{external-separability},\footnote{The degree of how community’s vertices are well separated from boundary vertices.} or both. The mutual property of all these methods (apart from \textit{Average degree}) is that they all degrade when the boundaries fade, that is, when the fraction of inter-connections arises.

Sixth, we showed that \textit{CMMAC} is a suitable solution for identifying malicious communities in an OSN, such as Socialbot Networks. Their fake users connect randomly to other normal users and then detach their internal edges~\cite{SocialBotNetworkBoshmaf2011}. However, we presume \textit{CMMAC} would be less effective in cases where the malicious communities present a ``more specific'' strategy of connecting to other communities, other than the ``dual-preferential'' attachment, such as connecting to vertices with similar attributes. We believe the described case will result in fewer ``unexpected'' edges, which in their turn, will contribute more ``false'' data for \textit{CMMAC}'s link-predictor. In the future, we intend to improve our Anomaly-Infused Community-Structured Random Network Generator by adding an ``attribute-oriented'' attachment functionality, to simulate such cases. To enable \textit{CMMAC} to handle such cases, we plan to reinforce its link-predictor with features that are based on attributes of vertices.

Finally, according to the real-world non-labeled networks results (see Section~\ref{subSection:Results__UnlabeledRealWorldNetworks}), we demonstrate \textit{CMMAC} can be applied to detect anomalous communities ``in the wild'' in different domains by ranking communities that presented abnormal behavior at the bottom (see Table~\ref{table:Results__UnlabledRankings}). The two non-labeled datasets we tested are a relatively small sample to test, hence, we intend to test \textit{CMMAC} on more real-world non-labeled datasets. While uncovering anomalous communities in ``native-network''~\footnote{Data that can intuitively be represented by a network.} data, such as Reddit's \textit{r/Place} project network, is a trivial task, the Wikipedia revisions network is an example of structuring non-trivial data into a network and utilizing \textit{CMMAC} to detect anomalies within it. We depicted articles as communities and the Wikipedians who edited them as vertices, and by utilizing \textit{CMMAC}, we uncovered articles containing anomalous revisions history. By generalizing this example, we believe \textit{CMMAC} can be utilized to detect anomalies in a variety of domains, in which there exists data that can be modeled as a community-structured network and the resulting network contains a certain extent of overlap between communities.

\section{Conclusion and Future Work}
\label{sec:ConclusionFutureWork}
The detection of anomalous communities in complex networks is becoming progressively prominent in our networked world. We present a novel generic method for detecting abnormal communities based solely on the co-membership of vertices to communities. Our approach is composed of graph theory notions and straightforward yet accurate machine-learning-based link-prediction techniques. In addition, we developed an algorithm that generates an overlapping community-structured random network to empower further research in the field.

We evaluated our method on 1,000 networks generated by us and on 1,000 networks sampled from Reddit’s comments dataset, where each contained tens of thousands of vertices and edges. We demonstrated our method succeeds in the scenarios where other known methods fail, specifically, when the anomalous communities are well disguised in the background, namely, they are sparse and heavily connected to other communities. We further demonstrated our method could detect anomalous communities in real-world networks by uncovering a violent subreddit and a collaboration-failing subreddit in the Reddit comments network and a Wikipedia article filled with inciting revisions.

Our open framework can be instantly utilized to gain insights into any data modeled as a community-structured network while providing a cost-effective practice that reduces a massive space of potential anomalies to a relatively small, threshold-dependent number of options to explore. Future directions could be to add more structural features, such as edge weights, and to add vertex attributes to be used as features to enhance the community membership prediction ability in specific domains.

\section{Availability}
\label{sec:Availability}

One of the main goals of this study is to facilitate future research in the field of anomalous community detection. Therefore, the code that implements our Anomaly-Infused Community-Structured Random Network Generator (see Section ~\ref{subSection:AnomalyInfusedCommunityStructuredRandomNetwork Generator}) and the labeled networks datasets we created (see Section~\ref{subsubsection:LabeledDatasets}) are open. In addition, we publish the framework that implements \textit{CMMAC}, including the evaluation process used for the study. The Network Generator and \textit{CMMAC} can be found on the project’s 
\href{https://github.com/lapidshay/GenericAnomalousCommunitiesDetection}{\color{blue}{GitHub Page}}, and the labeled networks' datasets can be found on the project’s 
\href{https://drive.google.com/drive/folders/14TQfwExIbhoHulBKe6DA7Cj9qf7Cw9xG?usp=sharing}{\color{blue}{shared directory}}.

\section*{Acknowledgements}
We would like to thank Sarah Ruddle for editing and proofreading this article to completion.

\printbibliography

\newpage

\begin{appendices}

\setcounter{figure}{0}  
\setcounter{table}{0}  
\setcounter{algorithm}{0}  

\makeatletter
\renewcommand{\thefigure}{S\@arabic\c@figure}
\renewcommand{\thetable}{S\@arabic\c@table}
\renewcommand{\thealgorithm}{S\@arabic\c@algorithm}

\makeatletter

\section{Anomaly-Infused Community-Structured Random Network Generator}
\label{sec:Appendix__AnomalyInfusedCommunityStructuredRandomNetworkGenerator}

The following section describes in detail our Anomaly-Infused Community-Structured Random Network Generator algorithm. The algorithm pseudo-code is given in Algorithm ~\ref{algorithm:Appendix__NetworkGenerator}.

The algorithm receives as input eight parameters, which can be divided into two groups of parameters:
\begin{itemize}
    \item Normal communities parameters: (1) Normal community random network generating algorithm (denoted $alg_{norm}$), (2) a map of normal communities and their desired sizes to be created (denoted $comm\_sizes_{norm}$), (3) arguments needed for the network generating algorithm (denoted $args_{norm}$), and (4) the ratio of vertices of each normal community to be connected to other communities (denoted $inter\_p_{norm}$).
    
    \item Anomalous communities' parameters: Same types of parameters as the normal communities parameters, but for anomalous communities; denoted (5) $alg_{anom}$, (6) $comm\_sizes_{anom}$, (7) $args_{anom}$, and (8) $inter\_p_{anom}$ respectively. We emphasize that $inter\_p_{anom}$ indicates the inter-connections fraction between anomalous communities and normal communities. 
\end{itemize}

The algorithm starts by creating an empty network (line 2) and an empty map to be populated by the network partitions (communities) (line 3). Then, for each of the tuples ($alg_{norm}$, $comm\_sizes_{norm}$, $args_{norm}$, $inter\_p_{norm}$) and ($alg_{anom}$, $comm\_sizes_{anom}$, $args_{anom}$, $inter\_p_{anom}$), the algorithm passes twice through the list of communities sizes $comm\_sizes$:

\begin{itemize}
    \item At the first pass it generates random subnetworks utilizing the network-generating algorithm $alg$, the arguments $args$ and the sizes are given by $comm\_sizes$ (line 7), merges the subnetworks to the main network $G$, that is, adding the newly created vertices and edges to the main network (lines 8-9), and updates the partition map of each subnetwork (community) to contain its vertices (line 10).
    
    \item At the second pass, it uses the $InterConnect$ procedure to connect each community to other normal communities (line 12).
\end{itemize}

Procedure $InterConnect$ receives as input two parameters: A set of vertices of the current community $V^c$, and a fraction that determines the number of inter-connections to create $inter\_p$. The procedure connects each newly created normal or anomalous community to other normal communities, using the following routine:

\begin{itemize}
    \item It first calculates the number of vertices in the given community that should be connected to other communities and randomly selects them (lines 15-16).
    
    \item For each of the selected vertices, preferentially chooses another community to connect to (line 17).
    
    \item Preferentially chooses a vertex to connect to in the chosen community, adds the created edge to the main network, and the connected vertex from the current community to the other community’s partition (lines 18-20), following the intuition that it was likely connected to a ``central'' vertex in the other community, thus, becoming a part of its community.
\end{itemize}

The functions of choosing a community to connect to, \textit{ChooseWeighted}, and the vertices to connect to, \textit{ChoosePreferentially}, are named differently to avoid ambiguity; however, they follow the same preferential concept, that is, choose randomly by a probability that correlates to communities' sizes or a vertices' degrees, respectively.

\begin{algorithm}[H]
\caption{\newline Anomaly-Infused Community-Structured Random Network Generator}
\label{algorithm:Appendix__NetworkGenerator}
\begin{algorithmic}[1]

\Procedure{GenerateRandomNetwork}{}
\Require $alg_{norm}, comm\_sizes_{norm}, args_{norm}, inter\_p_{norm}, alg_{anom}, comm\_sizes_{anom}, args_{anom}, inter\_p_{anom}$
\State $G\gets empty\: undirected\: network$
\State $Partitions\gets empty\: map$

\ForEach{$tuple$ in $[(alg_{norm}, comm\_sizes_{norm}, args_{norm}, inter\_p_{norm}),\newline \hspace*{5em}(alg_{anom}, comm\_sizes_{anom}, args_{anom}, inter\_p_{anom})]$}
    \State $alg, comm\_sizes, args, inter\_p \gets tuple$ \textcolor{gray}{\:\:\:// unpack tuple}
    \For{$c=1$ to $|comm\_sizes|$}
        \State $G^c=<V^c,E^c>\gets$ CreateNetwork$\big(alg(comm\_sizes[c], args)\big)$
        \State AddVerticesToNetwork($G, V^c$)
        \State AddEdgesToNetwork($G, E^c$)
        \State AddVerticesToPartition($c, V^c$)
    \EndFor
    \For{$c=1$ to $|comm\_sizes|$}
        \State InterConnect($V^c, inter\_p$)
    \EndFor
\EndFor
\State \textbf{return} $G, Partitions$
\EndProcedure
\item[]
\Procedure{InterConnect}{}
\Require $V^c, inter\_p$
    \For{$i=1$ to $\big\lfloor{|V^c|\times p \big\rfloor}$}
        \State $u\gets$ ChooseRandomVertex($V^c$)
        \State $otherComm\gets$ ChooseWeighted($comm\_sizes_{norm}$)
        \State $v\gets$ ChoosePreferentially($V^{otherComm}$)
        \State AddEdgesToNetwork($G, \{(u,v)\}$)
        \State AddVerticesToPartition($otherComm, \{v\}$)
    \EndFor
\EndProcedure

\end{algorithmic}
\end{algorithm}

\section{Selection of Network Generation Parameters}
\label{sec:Appendix__SelectionOfNetworkGenerationParameters}

To create the fully simulated networks (see Section~\ref{subSubSection:Datasets__SimulatedNetwork}) we utilized our Anomaly-Infused Community-Structured Random Network Generator (see Appendix~\ref{subSection:AnomalyInfusedCommunityStructuredRandomNetwork Generator}), which receives as input the parameters $alg_{norm}$, $comm\_sizes_{norm}$, $args_{norm}$, $inter\_p_{norm}$, $alg_{anom}$, $comm\_sizes_{anom}$, $args_{anom}$, and $inter\_p_{anom}$. To create the ``anomalous'' part of the anomaly-infused Reddit-based networks (see Section~\ref{subSubSection:Datasets__RealWorldNetworksWithArtificialAnomalies}) we used a partial functionality of our network generator, which requires only the input of the parameters $alg_{anom}$, $comm\_sizes_{anom}$, $args_{anom}$, and $inter\_p_{anom}$. This section describes the selection of the parameters.

The average degree distribution of the communities in Reddit’s network follows a power-law distribution. In particular, the mean average degree of the communities we sampled to create the networks equals 3.28. 
To imitate properties such as in Reddit’s network, we chose the following parameters to create the ``normal’’ part of the fully simulated network: (1) $alg_{norm}=$Barab\'asi-Albert~\cite{Barabasi1999} since it encapsulates both growth and preferential attachment, which are significant properties in real networks; (2) $comm\_sizes_{norm}$ were chosen by sampling communities’ sizes from the Reddit’s network; (3) $args_{norm}=1$, in this algorithm's case $m=1$,~\footnote{Each new vertex is attached preferentially by one edge to one of the existing vertices.} to produce a degree distribution similar to Reddit’s network; and (4) $inter\_p_{norm} = 0.075$, which we derived from the average percent of vertices that are part of cut-edges, in each of Reddit’s network’s communities.

To report reliable results and to study the strengths and weaknesses of our algorithm, we used a wide range of parameters’ values to create the ``anomalous’’ part of each of the networks: (1) $inter\_p_{anom}$ were set to $[0.05, 0.1, 0.15, 0.2, 0.25, 0.3, 0.35, 0.4]$; and (2) $comm\_sizes_{anom}$ were set to [$q_{0}$, $q_{0.1}$, $q_{0.25}$, $q_{0.5}$, random], where the names correspond the quantiles of the normal communities’ sizes distribution. To get exposed to the points where our method changes from underperforming to outperforming the baselines and to enable higher-resolution examination of them, we used two different value ranges for the $args_{anom}$ parameter.
Specifically, for the Reddit-based networks we set $args_{anom}$ to be the logarithmic scale $[0.05, 0.1, 0.2, 0.4, 0.8]$ and for the fully simulated networks we set $args_{anom}$ to be on a lower logarithmic scale $[0.01, 0.02, 0.04, 0.08, 0.16]$. The motivation for choosing the values is described as follows:

The expected average degree of a random community generated by the Barab\'asi-Albert algorithm equals $\mathop{\mathbb{E}}(\overline{k})=2\cdot m$. However, the ``dual-preferential'' inter-connectivity property of our Anomaly-Infused Community-Structured Random Network Generator, adds extra edges to each community, such that the expected average degree sums up to
$$\mathop{\mathbb{E}}(\overline{k})=2\cdot m + \dfrac{\overline{|V^c_{norm}|}\cdot inter\_p_{norm}}{|comm\_sizes_{norm}|}$$
where $m=1$, $\overline{|V^c_{norm}|}$ is the average normal community size, and equals 520, $inter\_p_{norm} = 0.075$, and $|comm\_sizes_{norm}|$ is the number of normal communities, which equals 110. The mean average degree of the normal communities in the generated networks results in $\overline{k}=2.35$.

The following concludes the reasons for selecting a lower logarithmic scale of values for the $args_{anom}$ parameter in the fully simulated networks:
The mean average degree of the normal communities in the Reddit-based networks is 47\% higher than the mean average degree of the normal communities in the fully simulated networks. This facilitates the \textit{internal-consistency}-based baselines to improve faster in the fully simulated networks than in Reddit-based networks. Moreover, we aimed to demonstrate where our method is also superior specifically to the \textit{avg. degree} method, which is the most rapidly affected method by the combination of community size and density.

\section{Train-Test Split Methodology}
\label{sec:Appendix__TrainTestSplitMethodology}

\textit{CMMAC} was developed to be a method for real-world uses; thus, we want to avoid consuming too many communities for the training phase at the expense of the extent of communities to test. Following the above utterance, we formulated our datasets such that the test sets contain the majority of the communities, and the train sets only contain enough communities to induce a sufficient number of edges for training.

Specifically, in the labeled network datasets described in Section~\ref{subsubsection:LabeledDatasets}, which we used for the evaluation, we used 20 communities for the train sets, which induced 18,000 positive and negative edges on average, and 100 communities in the test sets, which induced in 45,000 edges on average. In addition, the test sets contained ten anomalous communities out of the 100 communities.

In the real-world network datasets described in Section~\ref{subSubSection:UnlabeledRealWorldNetworks}, we had a trade-off between maximizing the potential number of anomalies to detect and filtering a portion of it to constrain overlap between communities, to be applicable for \textit{CMMAC}. We chose a compromise that yields an adequate degree of overlap as well as enough communities to test and then utilized most of the remaining data as follows: (1) 100 subreddit for training and 350 for testing in Reddit's r/Place dataset, (2) and 100 articles for training and 1,000 for testing in Wikipedia dataset.

\section{Quarry SQL query for obtaining Hebrew Wikipedia Revision Data}
\label{sec:Appendix__HebrewWikipediaSQL}

We utilized Quarry, an online public interface for running SQL queries against the Wikipedia database, to acquire
all revisions made to articles in the Hebrew Wikipedia between January $1^{st}$, 2016, and July $14^{th}$, 2020, using the query in Algorithm~\ref{algorithm:Appendix__AlgorithmHebWikiSQL}.

\begin{algorithm}[H]
\caption{\newline Quarry SQL query for obtaining Hebrew Wikipedia Revision Data}
\label{algorithm:Appendix__AlgorithmHebWikiSQL}
\begin{algorithmic}

\State USE hewiki\_p;

\State {SELECT
\State\hspace{\algorithmicindent}revision.rev\_id, revision.rev\_actor, revision.rev\_timestamp, revision.rev\_parent\_id, page.page\_title,
\State\hspace{\algorithmicindent}revision.rev\_minor\_edit,
revision.rev\_deleted}

\State FROM revision
\State JOIN page ON revision.rev\_page=page.page\_id
\State WHERE (
\State\hspace{\algorithmicindent}page.page\_namespace = 0
\State\hspace{\algorithmicindent}AND revision.rev\_timestamp BETWEEN 20160101000000 AND 2020715000000
\State\hspace{\algorithmicindent}page.page\_is\_redirect = 0
\State)

\end{algorithmic}
\end{algorithm}

\section{Results of Unlabeled Real-World Networks}
\label{sec:AppendixDResultsUnlabeled}

\subsection{Reddit's r/Place Network}
\label{subsec:RedditFullAlgoResults}

The following tables contain all the subreddits that were ranked at the three lowest rankings by each of the meta-features of \textit{CMMAC} (see Table~\ref{table:Appendix__TableRedditPlaceCMMACResults}) and by each of the other methods we utilized as baselines (see Table~\ref{table:Appendix__TableRedditPlaceOtherMethodsResults}).

\begin{table}[H]
    \begin{subtable}[t]{\columnwidth}
        \begin{tabular}{c|p{6cm}p{6cm}}
            \toprule
             Ranking & $EdgesNormalityMean$ &   $EdgesNormalitySTDV$ \\
            \midrule
                 348 &       starryknights &          cavestory \\
                 349 &           placesnek &     StrangerThings \\
                 350 &           cavestory &        necrodancer \\
            \bottomrule
        \end{tabular}
    \end{subtable}
    \hfill
    \begin{subtable}[t]{\columnwidth}
        \begin{tabular}{c|p{6cm}p{6cm}}
            \toprule
             Ranking & $PredictedEdgeLabelsMean$ & $PredictedEdgeLabelsSTDV$ \\
            \midrule
                 348 &          FloydVsVoid &     COMPLETEANARCHY \\
                 349 &             GreyBlob &          BlueCorner \\
                 350 &        theitalyplace &           cavestory \\
            \bottomrule
        \end{tabular}
    \end{subtable}
    \caption{Reddit's r/Place project subreddits ranked at the lowest three ranks by each \textit{CMMAC}'s meta-features\\(The table is split into two).}
    \label{table:Appendix__TableRedditPlaceCMMACResults}
\end{table}

\begin{table}[H]
    \begin{subtable}[t]{\columnwidth}
        \begin{tabular}{c|p{4cm}p{4cm}p{5cm}}
            \toprule
            Ranking & \textit{Average degree} & \textit{Cut ratio} & \textit{Conductance} \\
            \midrule
                 348 &   Philippines &             bdsm &            india \\
                 349 &     BABYMETAL &  hamiltonmusical &      Philippines \\
                 350 &   worldpowers &  britishcolumbia &  britishcolumbia \\
            \bottomrule
        \end{tabular}
    \end{subtable}
    \hfill
    \begin{subtable}[t]{\columnwidth}
        \begin{tabular}{c|p{4cm}p{4cm}p{5cm}}
            \toprule
             Ranking & \textit{Flake-ODF} & \textit{Average-ODF} & \textit{Unattributed-AMEN/ADENMN} \\
            \midrule
                 348 &  PuzzleAndDragons &             bdsm &             bdsm \\
                 349 &           paragon &  hamiltonmusical &  hamiltonmusical \\
                 350 &              bdsm &  britishcolumbia &  britishcolumbia \\
            \bottomrule
        \end{tabular}
    \end{subtable}
    \caption{Reddit's r/Place project subreddits ranked at the lowest three ranks by each of the methods we compare\\(The table is split into two).}
    \label{table:Appendix__TableRedditPlaceOtherMethodsResults}
\end{table}
\newpage
\subsection{Hebrew Wikipedia Revisions Network}
\label{subSec:HebWikiRevNetworkAppendixD}

The following tables contain all the articles that were ranked at the three lowest rankings by each of the meta-features of \textit{CMMAC} (see Table~\ref{table:Appendix__TableWikipediaCMMACResults}) and by each of the other methods we utilized as baselines (see Table~\ref{table:Appendix__TableWikipediaOtherMethodsResults}).

\begin{table}[H]
    \begin{subtable}[t]{\columnwidth}
        \begin{tabular}{c|p{6cm}p{6cm}}
            \toprule
             Ranking & $EdgesNormalityMean$ & $EdgesNormalitySTDV$ \\
            \midrule
                 998 &                    \<nw`h qyrl> &  \<h/sp`t mgpt hqwrwnh `l m`rkt h.hynwK> \\
                 999 &           \<.hsydwt qr`.t/snyP> &                    \<.hsydwt qr`.t/snyP> \\
                1000 &  \<'wpws hmgzyN lmwzyqh ql'syt> &                            \<'b' m.tplt> \\
            \bottomrule
        \end{tabular}
    \end{subtable}
    \hfill
    \begin{subtable}[t]{\columnwidth}
        \begin{tabular}{c|p{6cm}p{6cm}}
            \toprule
             Ranking & $PredictedEdgeLabelsMean$ & $PredictedEdgeLabelsSTDV$ \\
            \midrule
                 998 &  \<'wpws hmgzyN lmwzyqh ql'syt> &  \<'wpws hmgzyN lmwzyqh ql'syt> \\
                 999 &           \<.hsydwt qr`.t/snyP> &           \<.hsydwt qr`.t/snyP> \\
                1000 &                 \<rwny 'wslybN> &                 \<rwny 'wslybN> \\
            \bottomrule
        \end{tabular}
    \end{subtable}
    \caption{Hebrew Wikipedia revisions network's articles ranked at the lowest three ranks by each \textit{CMMAC}'s meta-features (The table is split into two).}
    \label{table:Appendix__TableWikipediaCMMACResults}
\end{table}

\begin{table}[H]
    \begin{subtable}[t]{\columnwidth}
        \begin{tabular}{c|p{4cm}p{4cm}p{5cm}}
            \toprule
            Ranking & \textit{Average degree} & \textit{Cut ratio} & \textit{Conductance} \\
            \midrule
                 998 &          \<mkby .hyph (kdwrgl)> &             \<`wrb> &  \<m'rq wyly'ms (/s.hqN snwqr)> \\
                 999 &                   \<'bnr byrwN> &           \<dwbyyM> &                 \<rwny 'wslybN> \\
                1000 &                   \<nyr qbr.ty> &  \<l/sbwr 't hqr.h> &                         \<mzlP> \\
            \bottomrule
        \end{tabular}
    \end{subtable}
    \hfill
    \begin{subtable}[t]{\columnwidth}
    
        \begin{tabular}{c|p{4cm}p{4cm}p{5cm}}
            \toprule
             Ranking & \textit{Flake-ODF} & \textit{Average-ODF} & \textit{Unattributed-AMEN/ADENMN} \\
            \midrule
                 998 &                    \<tpylyN> &             \<`wrb> &  \<l/sbwr 't hqr.h> \\
                 999 &                  \<trzh myy> &           \<dwbyyM> &             \<`wrb> \\
                1000 &                 \<t/s`h b'b> &  \<l/sbwr 't hqr.h> &           \<dwbyyM> \\
            \bottomrule
        \end{tabular}
    \end{subtable}
    \caption{Hebrew Wikipedia revisions network's articles ranked at the lowest three ranks by each of the methods we compare (The table is split into two).}
    \label{table:Appendix__TableWikipediaOtherMethodsResults}
\end{table}

\end{appendices}

\end{document}